\documentclass[prd,twocolumn,eqsecnum,superscriptaddress,showpacs,letterpaper,altaffilletter]{revtex4}
\usepackage{color}
\usepackage{times}
\usepackage{graphicx}
\usepackage{fancyhdr}
\usepackage{float}

\def\thercsid{\relax}
\def\rcsid#1{\def\next##1#1{\def\thercsid{##1}}\next}
\rcsid$Id: s1burstprd.tex,v 1.32 2003/12/07 03:43:35 ajw Exp $

\renewcommand{\today}{\number\day\space\ifcase\month\or
  January\or February\or March\or April\or May\or June\or
  July\or August\or September\or October\or November\or December\fi
  \space\number\year}

\def\rapprox{{\sim\kern-1em\raise 0.6ex\hbox{$>$}}}
\def\lapprox{{\sim\kern-1em\raise 0.6ex\hbox{$<$}}}
\def\mycaps#1{{\textsc{\small #1}}}
\newcommand\ligodoc{P030011}

\begin{document}

\title{First upper limits from LIGO on gravitational wave bursts \\
{\color{red} {\large LIGO-\ligodoc} }}

\begin{abstract}  
\vspace*{0.2in}
We report on a search for gravitational wave bursts
using data from the first science run of the LIGO detectors.
Our search focuses on bursts with durations ranging from 4 ms
to 100 ms, and with significant power in the LIGO
sensitivity band of 150 to 3000 Hz.
We bound the rate for such detected bursts at less than 
1.6 events per day at 90\%\ confidence level.
This result is interpreted in terms of the
detection efficiency for ad hoc waveforms (Gaussians
and sine-Gaussians) as a function of their 
root-sum-square strain $h_{rss}$;
typical sensitivities lie in the range 
$h_{rss} \sim 10^{-19} - 10^{-17}\,{\rm strain}/\sqrt{{\rm Hz}}$, 
depending on waveform.
We discuss improvements in the search method that will be applied
to future science data from LIGO and other gravitational
wave detectors.
\end{abstract}


\date[\relax]{ RCS \thercsid; compiled \today }

%
%
%
\newcommand*{\AG}{Albert-Einstein-Institut, Max-Planck-Institut f\"ur Gravitationsphysik, D-14476 Golm, Germany}
\affiliation{\AG}
\newcommand*{\AH}{Albert-Einstein-Institut, Max-Planck-Institut f\"ur Gravitationsphysik, D-30167 Hannover, Germany}
\affiliation{\AH}
\newcommand*{\AN}{Australian National University, Canberra, 0200, Australia}
\affiliation{\AN}
\newcommand*{\CH}{California Institute of Technology, Pasadena, CA  91125, USA}
\affiliation{\CH}
\newcommand*{\DO}{California State University Dominguez Hills, Carson, CA  90747, USA}
\affiliation{\DO}
\newcommand*{\CA}{Caltech-CaRT, Pasadena, CA  91125, USA}
\affiliation{\CA}
\newcommand*{\CU}{Cardiff University, Cardiff, CF2 3YB, United Kingdom}
\affiliation{\CU}
\newcommand*{\CL}{Carleton College, Northfield, MN  55057, USA}
\affiliation{\CL}
\newcommand*{\CO}{Cornell University, Ithaca, NY  14853, USA}
\affiliation{\CO}
\newcommand*{\FN}{Fermi National Accelerator Laboratory, Batavia, IL  60510, USA}
\affiliation{\FN}
\newcommand*{\HC}{Hobart and William Smith Colleges, Geneva, NY  14456, USA}
\affiliation{\HC}
\newcommand*{\IU}{Inter-University Centre for Astronomy  and Astrophysics, Pune - 411007, India}
\affiliation{\IU}
\newcommand*{\CT}{LIGO - California Institute of Technology, Pasadena, CA  91125, USA}
\affiliation{\CT}
\newcommand*{\LM}{LIGO - Massachusetts Institute of Technology, Cambridge, MA 02139, USA}
\affiliation{\LM}
\newcommand*{\LO}{LIGO Hanford Observatory, Richland, WA  99352, USA}
\affiliation{\LO}
\newcommand*{\LV}{LIGO Livingston Observatory, Livingston, LA  70754, USA}
\affiliation{\LV}
\newcommand*{\LU}{Louisiana State University, Baton Rouge, LA  70803, USA}
\affiliation{\LU}
\newcommand*{\LE}{Louisiana Tech University, Ruston, LA  71272, USA}
\affiliation{\LE}
\newcommand*{\LL}{Loyola University, New Orleans, LA 70118, USA}
\affiliation{\LL}
\newcommand*{\MP}{Max Planck Institut f\"ur Quantenoptik, D-85748, Garching, Germany}
\affiliation{\MP}
\newcommand*{\MS}{Moscow State University, Moscow, 119992, Russia}
\affiliation{\MS}
\newcommand*{\ND}{NASA/Goddard Space Flight Center, Greenbelt, MD  20771, USA}
\affiliation{\ND}
\newcommand*{\NA}{National Astronomical Observatory of Japan, Tokyo  181-8588, Japan}
\affiliation{\NA}
\newcommand*{\NO}{Northwestern University, Evanston, IL  60208, USA}
\affiliation{\NO}
\newcommand*{\SC}{Salish Kootenai College, Pablo, MT  59855, USA}
\affiliation{\SC}
\newcommand*{\SE}{Southeastern Louisiana University, Hammond, LA  70402, USA}
\affiliation{\SE}
\newcommand*{\SA}{Stanford University, Stanford, CA  94305, USA}
\affiliation{\SA}
\newcommand*{\SR}{Syracuse University, Syracuse, NY  13244, USA}
\affiliation{\SR}
\newcommand*{\PU}{The Pennsylvania State University, University Park, PA  16802, USA}
\affiliation{\PU}
\newcommand*{\TC}{The University of Texas at Brownsville and Texas Southmost College, Brownsville, TX  78520, USA}
\affiliation{\TC}
\newcommand*{\TR}{Trinity University, San Antonio, TX  78212, USA}
\affiliation{\TR}
\newcommand*{\HU}{Universit{\"a}t Hannover, D-30167 Hannover, Germany}
\affiliation{\HU}
\newcommand*{\BB}{Universitat de les Illes Balears, E-07071 Palma de Mallorca, Spain}
\affiliation{\BB}
\newcommand*{\BR}{University of Birmingham, Birmingham, B15 2TT, United Kingdom}
\affiliation{\BR}
\newcommand*{\FA}{University of Florida, Gainsville, FL  32611, USA}
\affiliation{\FA}
\newcommand*{\GU}{University of Glasgow, Glasgow, G12 8QQ, United Kingdom}
\affiliation{\GU}
\newcommand*{\MU}{University of Michigan, Ann Arbor, MI  48109, USA}
\affiliation{\MU}
\newcommand*{\OU}{University of Oregon, Eugene, OR  97403, USA}
\affiliation{\OU}
\newcommand*{\RO}{University of Rochester, Rochester, NY  14627, USA}
\affiliation{\RO}
\newcommand*{\UW}{University of Wisconsin-Milwaukee, Milwaukee, WI  53201, USA}
\affiliation{\UW}
\newcommand*{\WU}{Washington State University, Pullman, WA 99164, USA}
\affiliation{\WU}
\author{B.~Abbott}\affiliation{\CT}
\author{R.~Abbott}\affiliation{\LV}
\author{R.~Adhikari}\affiliation{\LM}
\author{A.~Ageev}\affiliation{\MS}\affiliation{\SR}
\author{B.~Allen}\affiliation{\UW}
\author{R.~Amin}\affiliation{\FA}
\author{S.~B.~Anderson}\affiliation{\CT}
\author{W.~G.~Anderson}\affiliation{\TC}
\author{M.~Araya}\affiliation{\CT}
\author{H.~Armandula}\affiliation{\CT}
% ...
 \collaboration{The LIGO Scientific Collaboration, http://www.ligo.org}
 \noaffiliation

\maketitle



\section{Introduction}\label{sec:intro}

Gravitational wave bursts are expected to be produced
from astrophysical sources such as stellar collapses,
the inspirals and mergers of compact binary star systems, 
the generators of gamma ray bursts, and other energetic phenomena.
Upper limits from
searches for gravitational wave bursts with resonant bar detectors
have recently been 
reported in \cite{ref:bars2000,ref:bars2001,ref:IGEC2002,ref:IGEC2003},
and results using interferometric detectors
are published in~\cite{ref:Forward,ref:UGMPQ}.
A new generation of detectors
based on laser interferometry has been constructed,
aiming for direct detection with broadband sensitivity.
These include the three LIGO detectors \cite{ref:LIGO}
described briefly in section~\ref{sec:detectors},
as well as the British-German GEO 600 detector \cite{ref:GEO,ref:GEO2}, 
the Japanese TAMA 300 detector \cite{ref:TAMA},
and the French-Italian VIRGO detector \cite{ref:VIRGO},
forming a worldwide network.
In the summer of 2002, all three LIGO detectors
were brought to their design optical configuration.
After a series of engineering runs,
the LIGO, GEO 600 and TAMA 300 detectors 
operated in coincident observation mode
for the first time (Science Run 1, or S1) 
for two weeks in August-September 2002.

Although the LIGO detectors were far from their design sensitivity,
the quality of the data was sufficiently high to exercise
the first generation of analysis procedures for various types
of gravitational wave searches, including searches for
chirp gravitational waves from
compact neutron-star binary inspirals \cite{ref:IUL},
quasi-monochromatic gravitational waves from pulsar J1939+2134 \cite{ref:PUL},
and broad-band stochastic background gravitational radiation \cite{ref:SUL}.
In all these analyses, a well-defined astrophysical model was assumed.
In this paper we report on a search (using the LIGO detectors only)
for unmodeled gravitational wave bursts
that might originate from supernovae in our galaxy,
mergers of binary stellar-mass systems, gamma ray burst engines,
or other energetic sources.
The waveforms of gravitational waves from such sources
are poorly known, so we employ data analysis algorithms which can,
in principle, identify bursts with a broad range of possible waveforms.

The first detection of gravitational wave bursts
requires stable, well understood detectors;
well-tested and robust data processing procedures;
and clearly defined criteria for establishing confidence
that no signal is of terrestrial origin.
None of these elements were firmly
in place as we began this first LIGO science run;
rather, this run provided the opportunity for us to
understand our detectors better,
exercise and hone our data processing procedures,
and build confidence in our ability to establish
detection of gravitational wave bursts in future science runs.
Therefore, the goal for this analysis is to produce an upper limit
on the rate for gravitational wave bursts,
even if a purely statistical procedure
suggests the presence of a signal above background.
It should also be noted that
the sensitivities of the three LIGO detectors during S1
were several orders of magnitude
worse than required for plausible detection
of bursts from astrophysical sources 
such as supernovae in our Milky Way galaxy
\cite{thorne87a}.

In this search we focus on short (4 ms to 100 ms) bursts
in the LIGO sensitivity band (roughly 150 to 3000 Hz),
with sufficiently high strain amplitude to be observed over the
detector noise. 
We make no other assumptions about the nature or origin of the burst.
We apply software algorithms to the LIGO detector data stream
to detect such bursts.
In order to suppress false signals
from fluctuations of the detector noise we require 
temporal coincidence of detected burst events in all
three LIGO detectors.
We estimate the rate of accidental coincidences by 
studying the number of time-shifted coincident burst events,
and look for a
statistically significant excess of coincident burst events
at zero time shift.
In light of the discussion in the previous paragraph,
our goal for the search presented here
is to set an upper limit on the rate of 
excess coincident bursts,
given the detectors' level of sensitivity during the S1 run.

In order to interpret our upper limit on the rate of burst events,
we evaluate the efficiency
of our search algorithms for the detection of simulated
bursts injected into the data streams, using
simple, well-defined waveforms (Gaussians and sine-Gaussians).
We obtain curves of triple-coincidence detection efficiency
as a function of gravitational waveform peak amplitude at the Earth,
averaged over source direction and incident wave (linear) polarization.
We then combine our gravitational wave burst rate limits
with these efficiency curves, yielding rate-versus-strength
regions that (for the waveforms that we have examined)
are excluded at the 90\%\ confidence level or higher.

The paper is organized as follows.
In section \ref{sec:detectors} we briefly describe the LIGO detector array
and the data obtained from the first science run,
with emphasis on those characteristics most relevant for a search
for short gravitational wave bursts.
In section  \ref{sec:S1} we briefly describe the S1 run.
In section  \ref{sec:preselection} we describe the data quality requirements
that were applied to the S1 data sample,
and present the subset of the data used for this search.
In section  \ref{sec:pipeline} we describe our data processing pipeline,
including the event trigger generation, event vetoes, and the
time coincidence requirement.
We present the results of two
independent pipelines, based on the burst detection
algorithms discussed in section~\ref{sec:ETGs}.
In section  \ref{sec:background} we estimate the background 
(accidental coincidence) event rate.
In section  \ref{sec:efficiency} we evaluate the 
efficiency for the detection of bursts modeled with simple
ad hoc waveforms, and compare that with expectations.
In section  \ref{sec:results} we present our limit on the
observed excess event rate.
We combine this with our efficiency curves as a function
of signal strength, excluding regions in the rate versus signal strength plane.
We also discuss the most significant systematic errors in these measurements.
We summarize these results in section \ref{sec:summary}.
Finally, we outline our plans to improve and expand
our search methodology using data
from subsequent observation runs.


\section{Detectors and Data Set}\label{sec:detectors}

\subsection{The LIGO detectors}\label{sec:LIGOdetectors}

All three LIGO detectors are orthogonal arm 
Michelson laser interferometers.
The LIGO Hanford Observatory 
operates two identically oriented interferometric detectors
which share a common vacuum envelope:
one having 4 km long measurement arms (referred to as H1), 
and one having 2 km long arms (H2).
The LIGO Livingston Observatory operates a single 4 km long detector (L1). 
The two observatories are approximately 3000 km
apart, corresponding to 10 ms of light travel time.
The detectors are approximately co-aligned, so that a gravitational wave
should appear with comparable signals at both sites.
The principles underlying these laser interferometer
gravitational wave detectors are discussed in \cite{ref:Saulson:1994}.
A more detailed description of
the LIGO detectors can be found in \cite{ref:LIGOS1instpaper}.

These detectors aim to detect gravitational waves
by interferometrically monitoring the relative separation of mirrors
which play the role of test masses,
responding to space-time distortions induced
by the waves as they traverse the detectors.
The effect of a quadrupolar gravitational wave 
is to produce a strain in space,
impinging upon the detector and
thus displacing the mirrors at the ends of the arms
by an amount proportional to the arm length.
For gravitational waves incident from directly overhead
or below,
and polarized along the arms of the detector,
the mirrors at the ends of the two arms experience
purely differential motion. 
Waves incident from non-optimal directions and/or polarizations
can also induce differential motion;
the ``antenna pattern'' is discussed in section~\ref{sec:direction}.

Each interferometer is illuminated with light from a 
Nd:YAG laser, operating at 1064 nm~\cite{Savage:1998}.
Before the light is launched into the interferometer, its frequency,
amplitude and direction are all stabilized, using a combination
of active and passive stabilization techniques~\cite{Kawamura97,Savage:1998}.
The light is sent through a beam splitter
towards both arms.
In each arm, a pair of mirrors (the ``input test mass''
and ``end test mass''),
separated by 2 km or 4 km,
form a Fabry Perot resonant optical cavity with a finesse of 
approximately 220.
Because the Michelson interferometer antisymmetric port 
is held at a dark fringe, and because the Fabry-Perot cavities are low-loss,
most of the light returning from the arms to the beam splitter 
nominally exits through the symmetric port of the beam splitter
back towards the laser.
A ``power recycling'' mirror returns it,
resonantly, to the interferometer
(forming a ``power recycling cavity'').
The average length of the arm cavities is used as a frequency reference for the
final stage of frequency stabilization~\cite{ref:LIGOS1instpaper}.
Differential arm cavity length changes 
result in a small amount
of light exiting the asymmetric port of the beam splitter;
this constitutes the gravitational wave signal.
The effect of the arm cavities and power
recycling is to increase the sensitivity of the interferometer
to gravitational wave signals.
The arm lengths and arm cavity finesse
are optimized to minimize various noise sources.

The mirrors of the interferometer \cite{ref:mirrors97,ref:mirrors99}
are suspended as pendulums \cite{ref:suspensions}.
Active and passive vibration isolation systems \cite{Giaime:1996}
are used to isolate them from seismic noise.
Various feedback 
control systems are used to keep the
multiple optical cavities tightly on resonance \cite{Fritschel:2001}
and well aligned \cite{Fritschel:1998},
and to keep the Michelson interferometer on a dark fringe.
The L1 detector also employed
feedforward control to compensate
for microseismic disturbances~\cite{Giaime:2003}.
When all length
degrees of freedom are under control and
the control systems are operating 
within their linear regime, the
interferometer is said to be ``in lock''.
During the first few minutes following the acquisition of a 
lock in any individual detector, the instrument
typically experiences excess noise due to the ringing down
of mechanical resonances in the mirror suspensions that were
excited by impulsive forces applied during
the lock acquisition procedure.
After 
allowing for these resonances to damp down, 
the detector is placed into ``science mode'';
the data collected in science mode are available
for gravitational wave searches.
Science mode continues until the interferometer
loses lock or becomes unstable for any reason.
The gravitational wave strain signal (referred to in this paper
as the gravitational wave data channel) is derived from the error
signal of the feedback loop used to control the differential
length of the interferometer arms. 
A 16 bit analog-to-digital converter
is used to digitize the (uncalibrated) strain signal at a rate of
16384 Hz.

To calibrate the error signal, the response to
a known differential arm strain is measured, 
and the frequency-dependent effect of the feedback loop
gain is measured and compensated for.
The laser wavelength and the amplitude of the mirror drive 
signal required to move the
interference pattern through a fixed number of fringes are used to
calibrate the absolute scale for strain.
The frequency response of the detector is determined
via periodic swept-sine excitations of the end test masses.
During detector operation, the calibration
is tracked by injecting continuous,
fixed-amplitude sinusoidal excitations into the
end test mass control systems,
and monitoring the amplitude
of these signals at the measurement (error) point.
The calibration procedure, and results, are described
in more detail in \cite{ref:calibration,ref:calib2}.

\subsection{The S1 run}\label{sec:S1}

By the summer of 2002,
all three LIGO detectors were operating reasonably stably
and with reasonable in-lock duty cycle.
As discussed below, the strain sensitivities of all
three detectors were far from their design goals,
but were nonetheless sensitive to gravitational wave bursts
from energetic events in our Galactic neighborhood.
The LIGO Laboratory decided that it was an appropriate time 
for the first Science Run, S1.

The S1 run consisted of a 408 hour continuous period
from August 23 through September 9 of 2002,
during which data were collected from all three
LIGO interferometric detectors.
The state of each of the detectors
and the quality of the data being logged was continuously monitored
through automated and manual procedures.
As discussed above,
in order to be sensitive to gravitational waves, 
the detectors must be in science mode.
Environmental disturbances and various instrumental
instabilities make it impossible to maintain lock at all times,
reducing the effective observation time of the run.
During S1, the science mode
duty cycles of the three detectors were
41.7\%\ for L1, 57.6\%\ for H1, and 73.1\%\ for H2.
The burst search reported here makes use of the data when all
three detectors were in science mode simultaneously, comprising
95.7 hours, or 23.4\%\ duty cycle.

\begin{figure}
\includegraphics[width=0.95\linewidth]{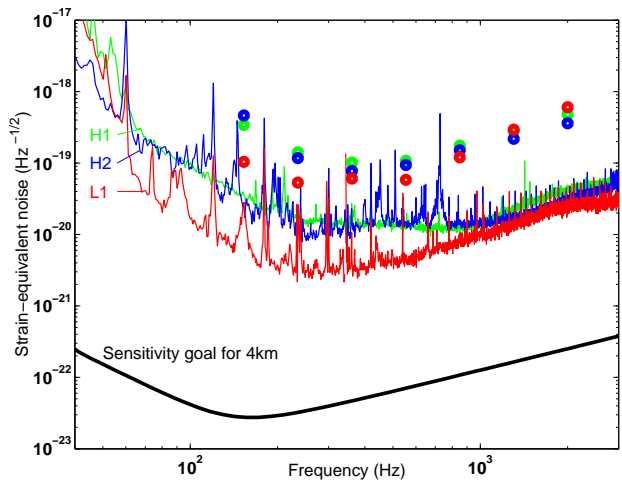}
\caption{Typical sensitivities of the three LIGO detectors during
the S1 data run, in terms of equivalent strain noise amplitude density.
The points are the root-sum-square strain ($h_{rss}$)
of sine-Gaussian bursts
for which our \mycaps{TFCLUSTERS}
analysis pipeline is 50\%\ efficient, as reported in 
section \ref{sec:simulations}.
}\label{fig:strain}
\end{figure}

The strain sensitivity of the LIGO detectors is a 
strong function of gravitational wave frequency.
In this analysis, we focus on a ``detection band'' 
of best strain sensitivity, from 150 to 3000 Hz.
Figure \ref{fig:strain} shows amplitude spectra of strain-equivalent
noise, typical of the three LIGO detectors during
the S1 run. The LIGO design strain sensitivity
is also indicated for comparison. 
The differences among the three
spectra reflect differences in the operating parameters
and hardware implementations of the three instruments;
they are in various stages of reaching the final design
configuration. All detectors operated during
S1 at lower effective laser power levels
than the eventual level of 6 W at the interferometer input.
Other major
differences between the S1 state and the final configuration
were partially implemented laser frequency and
amplitude stabilization systems and partially implemented
alignment control systems.
Because of these conditions, the strain sensitivities of the
three detectors were far from the design sensitivity
(see Figure \ref{fig:strain}).

\subsection{Data preselection}\label{sec:preselection}

The data processing pipeline described in section \ref{sec:pipeline}
makes use of many adjustable parameters that can be tuned
to optimize the search effectiveness. 
We performed these optimizations on a subset of the S1 data that
was reserved exclusively for the purpose, and then not used further in
the generation of scientific results.  We called this reserved data
set the ``playground'' data set.  It was chosen to be about 10\% of the
total available triple coincidence data. The choice of which data to
include was made by hand, to include as much variety of data quality
as possible.  The same playground data set was used for both the burst
search and the search for inspiralling binary neutron stars \cite{ref:IUL}. 
This tuning procedure is described in section \ref{sec:pipeline}.
Further, the data processing pipeline analyzed triple-coincidence data
in six-minute stretches, for convenience in data handling.
Lock stretches that were less than
six minutes long, or data in the last $< 6$ minutes of a longer lock stretch,
were excluded from further analysis.
After exclusion of the playground data and these lock stretch boundaries, 
80.8 hours of triple-coincidence data remain.

Much effort has gone into improving the stationarity of the
statistical properties of the detector noise, and understanding
the noise fluctuations. However, both the detectors' responses,
and their noise levels, were far from stationary, 
largely because the control systems 
were not yet completely implemented. In order
to ensure that the data used for this burst search are of the
highest available quality, we excluded locked stretches in
which the noise in the gravitational wave channel exceeded a
pre-determined threshold.  The band-limited root-mean-square (BLRMS)
noise power in the gravitational
wave channel was monitored continuously in four
bands (320 -- 400 Hz, 400 -- 600 Hz, 600 -- 1600 Hz, and
1600 -- 3000 Hz). Whenever
the BLRMS over a six-minute interval for any detector in any of these bands
exceeded a threshold of 3 times the 
68th percentile level for the entire run (10 times for the 320--400 Hz band),
the data from that six-minute period were
excluded from further analysis. 
A total of 54.6 hours of triple-coincidence
data remains after this ``BLRMS cut''.
A sufficiently strong
gravitational wave burst could trigger the BLRMS cut and thereby
prevent its own detection; the required amplitude is calculated
in section~\ref{sec:simulations}.

As discussed in section~\ref{sec:LIGOdetectors},
the response of the detectors to gravitational
waves was tracked by injecting sinusoidal calibration excitations into the
end test mass control systems. Due to technical difficulties, these
calibration lines were 
not reliable or available during some data taking periods.
In order to ensure that all the data
used in this search represent observations from detectors with
well-understood response, data that show no, or anomalously
low, calibration lines were excluded from further analysis (the
``calibration cut''), leaving 35.5 hours of triple-coincidence
data remaining. This is the final data sample used to search
for gravitational wave bursts.



\section{The data processing pipeline}\label{sec:pipeline}

In the analysis presented here, the purpose of the data processing
pipeline is to identify candidate gravitational wave events
in the data from all three detectors in coincidence.
In this section,
we discuss the procedures and algorithms used to identify
coincident burst event candidates, the tuning
of the most important parameters, and 
the procedures used to estimate the accidental 
coincident burst event rate.
The entire analysis procedure, parameter tuning, event
property estimation, and all other optimizations were developed
using the playground data (section~\ref{sec:preselection}),
and frozen before applying the analysis to the full S1 data set.
In the process of analyzing the full data set, 
it became clear that many of the procedures and 
tunings were less than optimal, for a variety of reasons.
We present the results of this first
analysis in this paper, and intend to
apply improved methods and optimizations
(see section \ref{sec:future}) to the analysis
of future data sets (which will have much greater sensitivity
to gravitational wave bursts).

\subsection{Pipeline overview}

\begin{figure*}[!t]
\includegraphics[width=0.95\linewidth]{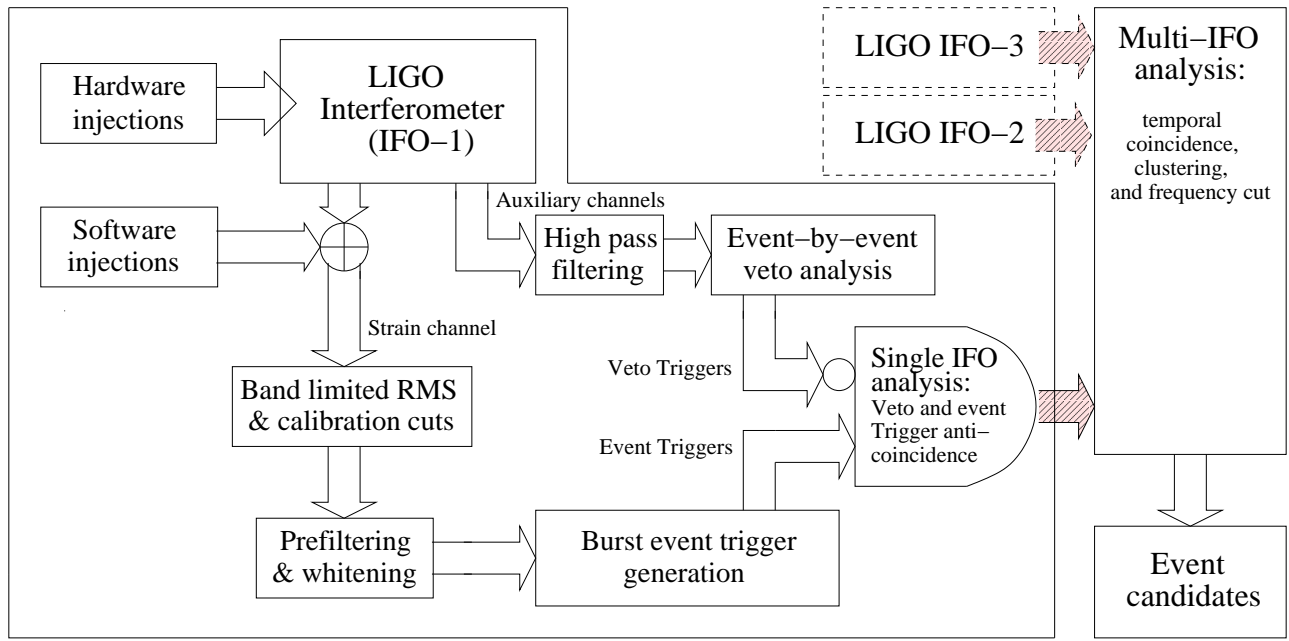}
\caption{Schematic outline of the S1 bursts analysis pipeline.
}
\label{fig:pipeline}
\end{figure*}

Figure \ref{fig:pipeline} shows, in graphical form, the data
processing pipeline used in this analysis. 
Most of the figure is used to schematically illustrate various steps in
the pipeline of one of the interferometric
detectors (H1, L1, or H2, generically referred to here as IFO-1). 
The analysis pipelines
of the other two IFOs (IFO-2 and IFO-3) are not shown in detail because
they are identical to the first.
The first step in the pipeline 
(``Band limited RMS \&\ calibration cuts'')
validates the strain channel data used in the analysis;
only validated data (section \ref{sec:preselection})
taken at times when all three detectors were operating 
simultaneously in science mode are used in this analysis.
This step establishes
the accumulated observation time, or livetime, for the analysis.

The next steps in the pipeline (``Prefiltering \&\ whitening'')
take as input the raw gravitational wave channel
data from each detector, and prefilter the data stream
(section \ref{sec:prefiltering}).
The following step (``Burst event trigger generation'')
searches for bursts in the filtered data stream
using two different burst detection algorithms
(section \ref{sec:ETGs}),
resulting in a set of \emph{event triggers} at each detector.
All data were processed in non-overlapping segments that were
six minutes long.

Our pipeline allows for the elimination of 
event triggers that are coincident in time
with anomalous events in 
auxiliary channels that monitor the detector and the environment
(see ``Auxiliary channels'' path and ``Single IFO analysis''
gate in Fig.~\ref{fig:pipeline}).
The consideration of these potential vetoes will be
described in section \ref{sec:vetoes}.

Real gravitational wave bursts will cause a nearly
simultaneous response in all three detectors,
so in the next step
we require temporal coincidence of single-detector event triggers
(``Multi-IFO analysis'' block in Fig.~\ref{fig:pipeline}).
We select as ``Event candidates'' only those combinations
of single-detector event triggers 
that are consistent with originating from 
a single plane gravitational wave burst incident
on the detector array (section \ref{sec:coincidence}).

Event triggers from the single-detector steps
of our pipeline are mostly due to detector noise.
The independence of noise triggers at the two LIGO sites is an
important assumption in this analysis.  The largest sources
of noise above 100 Hz are generated internal to the detectors
and are thus uncorrelated.  Environmental disturbances can
produce bursts of noise that cause triggers, and these can be
coincident between the two sites within the 
$\pm$10 ms gravitational
wave travel time if they propagate electromagnetically. 
Auxiliary sensors ({\it e.g.}, magnetometers) monitor the environment
and their inter-site correlations have been studied.
Inter-site disturbances
are calculated not to be important until the detectors are close to
design sensitivity, and our studies of S1 data have not
found evidence for coincident noise bursts even for H1-H2
where the coincident location increases the susceptibility to
environmental disturbances.
There does exist some coherence between the 
gravitational wave channels from the three detectors
at certain frequencies~\cite{ref:SUL},
but there is no evidence that this contributes to 
coincident noise bursts.

Gravitational wave burst events detected at the two LIGO sites
will be correlated in time.
We can evaluate the mean
rate of background events by measuring the mean rate of events that
pass our coincident step after we have 
artificially shifted in time all the event
triggers identified in one of the detectors, for example, L1.
This background rate estimation
is described in section \ref{sec:background}.

Finally, to determine the efficiency of the data processing pipeline to
the detection of strain events incident on the detector array we add
simulated events, 
of varying waveform and amplitude, to the input data stream and
measure the fraction identified as event triggers in each detector.
Knowing the detectors' sensitivity to gravitational waves
incident from different directions we can combine the results of these
simulations to determine the mean efficiency for detection of
gravitational wave burst events incident on the detector array.
The efficiency determination is described in section \ref{sec:efficiency}.



\subsection{Prefiltering}\label{sec:prefiltering}

The event trigger generators we employ are designed
to process data with a white noise spectrum
(constant power spectral density as a function of frequency).
The raw gravitational wave data 
from all three detectors during S1
are strongly colored, consisting essentially of randomly fluctuating noise
with a strongly frequency-dependent power spectrum. 
These data can be converted to a noise equivalent strain signal
through a response function which is also strongly
frequency-dependent, and which is determined through the
calibration procedure described in section \ref{sec:detectors}.
The noise also contains unwanted features
such as spectral lines associated with interference from the 60 Hz power mains,
mechanical resonances in the detector components, and other
imperfections.

For the analysis presented here, the data from the
gravitational wave data channel are passed through a linear
filter, consisting of a 6-th order Butterworth
high-pass filter with 150 Hz
cut-off frequency to suppress large noise fluctuations
which were apparent at lower frequencies, 
and a whitening filter to flatten the noise spectrum at frequencies
above 150 Hz.
Because of the high-pass filtering, we are insensitive
to Fourier components of a gravitational wave burst 
below 150 Hz.
The whitening filters are determined using data taken 
just prior to the S1 run, and are different for each of the
three detectors. No attempt has been made to incorporate
the variation of the noise power with time, or to otherwise
optimize the filtering. Further, no attempt has been made
to remove spectral lines from the data stream or 
suppress their effect on the event trigger identification.
It is likely that such pre-processing will be necessary
for future, more sensitive searches with LIGO data~\cite{ref:meaculpa}.

The impulse response of the prefilter used for this analysis 
has a strong ringing, extending to 40 ms. 
As a result of this ringing, 
the event trigger generation algorithms 
recognize an impulsive event in the strain channel as
a cluster of events over a long period of time compared to both the
sample rate and the light travel time between the detectors. 
This
has important consequences for the event trigger time resolution
and the time coincidence of event triggers generated in different
detectors, as described in section \ref{sec:coincidence} 
below~\cite{ref:meaculpa}. 



\subsection{Event Trigger Generation}\label{sec:ETGs}

We use two different techniques to identify event triggers from
the prefiltered gravitational wave data channel at each detector. 
One technique, which we refer to as \mycaps{SLOPE}, is based on 
Refs.~\cite{arnaud99a,pradier01a}. 
The second technique, which we refer to
as \mycaps{TFCLUSTERS}, is described in
Refs.~\cite{sylvestre02a,sylvestre02b}.  
\mycaps{SLOPE} and 
\mycaps{TFCLUSTERS} are two different approaches to
identifying and selecting infrequent transient events 
that do not share the statistical characteristics of detector noise 
and thus might be of gravitational wave origin. 
These algorithms are implemented within the LIGO Data Analysis
System (LDAS~\cite{ref:LDASLAL}) environment.

The following discussion of the 
\mycaps{SLOPE} and \mycaps{TFCLUSTERS} event trigger generators
describes and specifies the parameters that can be adjusted in order
to optimize the performance of the algorithms. 
Some of the parameters can be established without reference
to the data, since they effectively determine the
response of the algorithm to the 
duration (4 ms to 100 ms)
and frequency band (150 to 3000 Hz)
characteristics of the bursts
that are targeted in this search.
Others have been optimized using the playground data defined
in section \ref{sec:preselection}. 
It is assumed that no (or very few)
real gravitational wave bursts
were present in the playground sample.
All parameters were fixed 
prior to the processing of the full data set, 
in order to minimize the chance of bias 
in event trigger generation.

The parameter optimization, especially the choice of thresholds,
is guided by competing demands.
Lower thresholds on excess power or amplitude variations
result in higher rates of event triggers caused by noise fluctuations,
but also result in higher sensitivity to gravitational wave bursts.
The criterion we adopted consists of minimizing 
the upper limit 
for a suite of simulated gravitational wave bursts,
described in section \ref{sec:efficiency}. 
This minimization was applied to
the playground data set 
where no triple coincidence event was found after the thresholds 
were fixed. This was consistent with a goal of 
obtaining a total number of 
accidental coincident triggers of order unity, when extrapolated to 
the remaining 90\%\ of the full S1 data set.

Nevertheless,
the playground data did not adequately represent
the full S1 data set, and a variety of additional
effects (including the ringing in the prefiltering,
as discussed in section \ref{sec:prefiltering})
resulted in imperfect optimization of the 
data processing pipeline for both event trigger generators \cite{ref:meaculpa}.
Therefore, the resulting number of 
estimated accidental coincidence events
was somewhat larger than one, as discussed in 
section \ref{sec:bound} and Table \ref{tbl:S1raw}.

\subsubsection{\mycaps{SLOPE}}\label{sec:SLOPE}

The \mycaps{SLOPE} algorithm identifies candidate gravitational wave
bursts via a threshold on the output of a linear filter
applied to the prefiltered gravitational wave data in the time domain.
We choose a filter that is essentially a differentiator (in time),
and trigger on a slope in the data stream which is 
(statistically) inconsistent with expectations from 
white Gaussian noise.
The \mycaps{SLOPE} algorithm is
most sensitive when the detector noise in the strain channel 
is whitened.

The parameters of the \mycaps{SLOPE} filter have been tuned so that
its highest sensitivity is for bursts in which the signal 
amplitude is increasing linearly with time for ten data samples
($10\times\mathrm{61 \mu sec}$). 
The response of the filter to sine waves rises
with frequency from zero at DC, reaching its first and highest
maximum at 1.1 kHz. Above this frequency, the response of the 
filter falls off, passing through several zeros and secondary maxima.
Its 3 dB bandwidth is about 1.4 kHz~\cite{ref:meaculpa}.

The filter output
is searched for extrema
indicating the presence of bursts. The peak search algorithm compares
each successive filter output value with a threshold. If a 
filter output value
is found to exceed the threshold, then that point and some number of
output filter value after the first point exceeding the threshold are
further analyzed. For the analysis considered here, 49 output filter values
including the point that passed threshold are examined, a time interval
of 3.0 ms. The output filter value
having the highest value in this time
interval generates a single trigger. The amplitude of the trigger and
time of the trigger are written to a trigger database.
For this analysis, the threshold was fixed and did not adapt
to changing noise levels~\cite{ref:meaculpa}.



\subsubsection{\mycaps{TFCLUSTERS}}\label{sec:TFCLUSTERS}

The \mycaps{TFCLUSTERS} event trigger generator is a detection algorithm which
identifies connected regions ({\em clusters}) in a time-frequency
plane where the power is not consistent with the expectations for 
stationary, colored Gaussian noise.  
The \mycaps{TFCLUSTERS} algorithm is described in
detail in \cite{sylvestre02b}, and various aspects of its
implementation for real data are discussed in \cite{sylvestre02a}.
The implementation of \mycaps{TFCLUSTERS} used for our analysis is described
below.  

The data from a six minute long segment are first prefiltered
as described in section \ref{sec:prefiltering}.
A time-frequency spectrogram is constructed from 2880
periodograms calculated from 125 ms long non-overlapping
subsegments of the six minute long segment~\cite{ref:meaculpa}. 

A first level of threshold is applied to the spectrogram,
resulting in a high-contrast pixelization.
2880 different measurements of the power are
available for every frequency band of the spectrogram. 
Processing one frequency band at a time, 
the power measurements are fit with a
Rice distribution~\cite{ref:Rice}.
Given this fit to the data, the
Rice distribution is integrated from a power $\eta$ to infinity, and
$\eta$ is varied until the integral is equal to a certain pre-defined
fraction $p$. All the pixels of the spectrogram with power larger than
$\eta$ are then labeled as {\em black pixels}, while pixels below the
threshold are labeled as white pixels. The procedure was repeated for all
the frequency bins in the spectrogram. The number $p$ is called the {\em
black pixel probability}: in the absence of signals, any pixel in the
spectrogram has, to a good approximation, an equal and independent
probability $p$ of being black, in each frequency band.
Because of this procedure, the effective threshold for black pixels
varies in response to changing detector noise levels;
the threshold is ``adaptive'',
as opposed to the fixed threshold employed in the SLOPE algorithm.

The black pixels are then clustered, to look for bursts of excess 
power in a limited region of the time-frequency plane.
Two levels of clustering are used by \mycaps{TFCLUSTERS},
based on a study of simulated bursts with varying waveforms.
First, a cluster is defined as the set of all
black pixels which has at least one black nearest neighbor (i.e., was
touching a black pixel by an ``edge'') in the set. All clusters
containing at least five pixels are declared significant in this
analysis. 
Second,
clusters which are not significant according to the
latter criterion are paired together. If the clusters in a pair are
closer to each other in the time-frequency plane than a certain
distance threshold, the pair of clusters is declared significant. 
  
Clusters satisfying the first clustering condition
on the raw size of a cluster are counted as event triggers.
For clusters satisfying the second clustering condition, {\em generalized}
clusters are formed by linking all the clusters which satisfy the
distance thresholds, and these generalized clusters
are counted as event triggers.
For each event trigger,
the time and frequency intervals over which
the cluster extends,
the total amount
of power in the cluster, and the number of pixels it contains,
are stored in a database.
The total power in each cluster 
is a measure of the signal-to-noise ratio for the burst event.
It is calculated without reference to the response of the detector
to gravational wave bursts, so its relationship to the
strength of the burst depends on the detector and frequency band.

The black pixel probability $p$ is tuned as described above.  The values are
different for the three different detectors and vary from 0.02 to 0.05.  
The total power in the cluster is required to exceed a pre-determined
threshold in post-processing;
this is effectively a cut on the signal-to-noise ratio for the burst event.
The threshold on the power is the same for all three detectors,
in order to obtain rates for false (noise) triggers which are roughly
the same for all three detectors.

\subsection{\mycaps{SLOPE} and
  \mycaps{TFCLUSTERS} event triggers}\label{sec:triggers}

\begin{figure}
\includegraphics[width=0.95\linewidth]{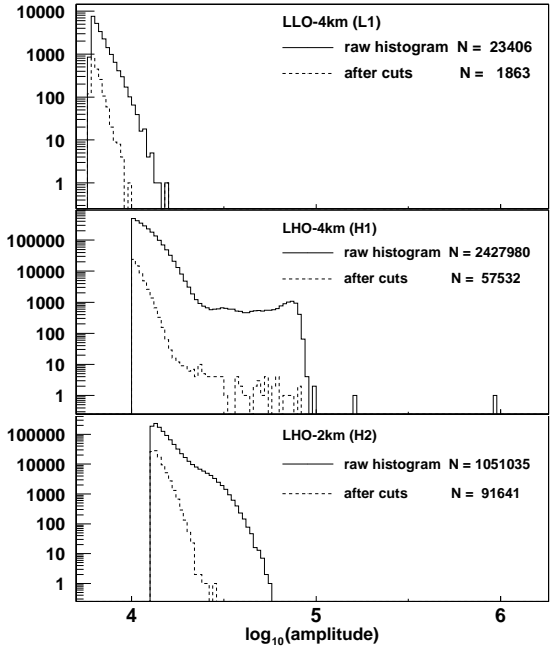}
\caption{Histogram of \mycaps{SLOPE} event triggers from the 
three LIGO detectors, before and after the BLRMS
and calibration cuts.}\label{fig:slopeHist} 
\end{figure}

\begin{figure}
\includegraphics[width=0.95\linewidth]{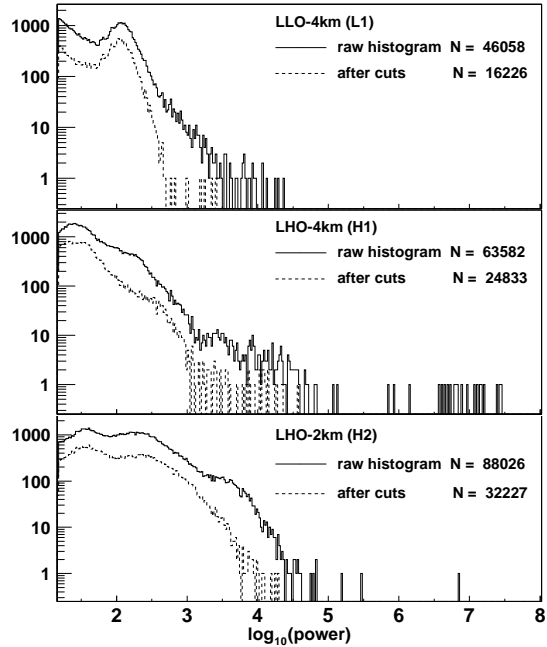}
\caption{Histogram of \mycaps{TFCLUSTERS} event triggers from the
three LIGO detectors, before and after the BLRMS
and calibration cuts.}\label{fig:tfcHist} 
\end{figure}

Figs.~\ref{fig:slopeHist} and \ref{fig:tfcHist} show histograms of
\mycaps{SLOPE} and 
\mycaps{TFCLUSTERS} event triggers before and after the application of the
BLRMS and calibration cuts described in section \ref{sec:preselection}.
The horizontal axis in these histograms is a measure of the 
amplitude or power of the excess signal identified
by the \mycaps{SLOPE} or \mycaps{TFCLUSTERS} algorithms, respectively.
These measures are indeed proportional to the true
amplitude or power of a detected gravitational wave burst,
as demonstrated in section \ref{sec:efficiency}.
However, no information about the detectors' calibrated
response functions is used in forming these measures,
so the proportionality constant is different
for different waveforms, detectors, and data epochs
(and is taken into account in the evaluation of the detection
efficiency, section \ref{sec:efficiency}).
The lower limits on the horizontal axis in these histograms correspond
to the threshold applied to that event trigger
for input into the next step in the data processing pipeline
(triple coincidence).

\subsection{Auxiliary Channel Vetoes}\label{sec:vetoes}

Environmental disturbances and detector instabilities
could also produce event triggers.
We collect data in a large number
of auxiliary channels which monitor the
detector and the environment, 
in order to look for time-coincident bursts and
thus form vetoes for such false triggers.
Our pipeline has the capability to search
for such bursts in auxiliary channels, and veto an event trigger
if it is time-coincident with such a burst.
Engineering runs performed prior to
the S1 run indicated that such vetoes could be very efficacious,
reducing the rate of false event triggers with minimal 
loss of livetime,
due to clearly identifiable instabilities in the detectors.
However, once these instabilities were identified, they were
eliminated through improved instrumentation,
resulting in much improved stability during S1.
After careful study,
no vetoing criteria using auxiliary interferometer and
physical environment monitor channels are found to be 
especially efficacious in the S1 data, for this burst search.
The most promising vetoing channels in the S1 data are interferometer 
sensors that are 
closely related to the gravitational wave channel.
While we investigated a number
of such channels and methods for identifying veto criteria,
in the end we concluded that further study was needed before any of these
could safely be used to exclude data from analysis. 
Further, employing the identified vetoes would have made
a negligible difference in the results of this analysis.
Thus, in this analysis, we apply no vetoes based on auxiliary channels.



\subsection{Coincidence}\label{sec:coincidence}

The final stage of our data processing
pipeline brings together the event triggers generated by a particular
event trigger generator 
(either \mycaps{SLOPE} or \mycaps{TFCLUSTERS}) and assembles a
smaller list of coincident event trigger triplets. 
Each triplet consists of an event trigger from each
detector that occur
within an interval consistent with their origin in a single
gravitational wave burst.  These triplets are the event candidates
that form the basis for our determination of 
bounds on the rate of gravitational wave bursts incident on
the Earth.

Temporal coincidence is the most obvious application
of coincidence for selection of gravitational wave events and exclusion of
noise events.  The LIGO detectors are approximately co-aligned and
co-planar. As a result, they all sense approximately the same polarization of
any incident gravitational wave. 
Correspondingly, all estimated parameters of the burst (such as strain
amplitude and frequency band) should be, up to uncertainties in
the estimation, the same for all three detectors (after accounting
for the differences in the detectors' sensitivities).
In the
analysis presented here we require temporal coincidence (to an
appropriate precision) for both the \mycaps{SLOPE} and the \mycaps{TFCLUSTERS}
pipelines.  Additionally, \mycaps{TFCLUSTERS} events are also
characterized by frequency information;
we require consistency between the frequency bands
in a coincident triplet (section~\ref{sec:freqCut}).
No attempt is made to require coincident event triggers to have
consistent amplitudes and waveforms~\cite{ref:meaculpa}.

In the remainder of this section we describe in greater detail the
elements of the data processing pipeline coincidence step. 

\subsubsection{Temporal coincidence}\label{sec:temporalCoinc}

Gravitational waves arrive at the Earth as plane waves. 
Since gravitational waves are assumed to
propagate at the speed of light,
the interval between event triggers in the different detectors should
be no more than the greater of the light propagation time between the
detectors and the uncertainty in the arrival time determination of a
prototypical burst associated with the event trigger generator.  
Different timing uncertainties are associated with different event
trigger generators. Correspondingly, we use different window durations
for \mycaps{SLOPE} and \mycaps{TFCLUSTERS}. 
Given a window, we compare the start times of the event triggers generated
in each of the three detectors. We form an \emph{event trigger
triplet}, or triplet for short, from all combinations of H1, H2 and
L1 events whose start times all lie within the window
duration.

As described in section \ref{sec:prefiltering}
the input to the event trigger generators is processed through
a high-pass filter that rang strongly. As a result of this ringing,
impulsive events lead to a train of multiple \mycaps{SLOPE} triggers,
with a total duration of 
approximately 40~ms. We add 10~ms to this, corresponding to the
light travel time between detectors, to determine a 50~ms window for
temporal coincidence of \mycaps{SLOPE} events~\cite{ref:meaculpa}.

As described in section \ref{sec:TFCLUSTERS},
\mycaps{TFCLUSTERS} was tuned to a natural time
resolution of 125~ms, much larger than the light travel time
between the detectors. 
On the basis of studies which indicated a larger range
of trigger time differences for simulated signals, we
expanded this and use a 500~ms window to determine triplets of
temporally coincident \mycaps{TFCLUSTERS} events~\cite{ref:meaculpa}.

\subsubsection{Clustering}\label{sec:clustering}

The next step in the multiple-detector coincidence analysis is to 
\emph{cluster} the events from each detector
(this is unrelated to the pixel clustering that forms the heart of the 
\mycaps{TFCLUSTERS} event trigger generation,
section~\ref{sec:TFCLUSTERS}).
Both the \mycaps{TFCLUSTERS} and the \mycaps{SLOPE} event trigger 
generators often associate several event triggers with the same 
``burst'' feature.  For instance, the ringing of a 1 ms Gaussian due to 
the detector response and the prefiltering of the data 
(section~\ref{sec:prefiltering}) can produce several closely spaced 
event triggers.  \mycaps{TFCLUSTERS} often associates multiple triggers 
with the same broadband event, all with the same start time but 
different frequency.
Since we are interested in the identification of time intervals where 
``something unusual'' has happened simultaneously at multiple 
detectors, we want to cluster these sets of closely spaced events.

Clustering takes place only after the time coincidence step.
We require a minimum
separation in time between distinct
coincident trigger triplets, of 0.5 seconds;
triplets that are separated in time by less than this amount
are clustered together into one clustered event triplet
(event candidate).
The choice of the clustering window is based on 
the study of noise triggers 
and simulated bursts (section~\ref{sec:simulations}).
In the \mycaps{TFCLUSTERS} pipeline,
0.5 seconds is the width of the coincidence 
window between triggers from the three detectors.
In the \mycaps{SLOPE} pipeline, 
the coincidence window of 50 ms is too 
small a separation to avoid ambiguities in the definition of clusters 
and in the event counting, so we use 0.5 seconds for consistency with 
the \mycaps{TFCLUSTERS} pipeline \cite{ref:meaculpa}.

All triggers in the cluster are assumed to originate from one burst event.
Guided by simulation studies (section~\ref{sec:simulations}), the start 
time, frequency band, and amplitude or power of the event is taken to be 
that of the trigger with the largest amplitude or power in the cluster.

\subsubsection{\mycaps{TFCLUSTERS} frequency cut}\label{sec:freqCut}

For TFLUSTERS we apply one more criterion in the coincidence step of the
pipeline. A triplet of event triggers
that arises from a
single gravitational wave burst incident on all the detectors
should have consistent values for the estimated parameters of the burst.
\mycaps{TFCLUSTERS} characterizes each burst event trigger by its bandwidth:
the low and high frequency bound $(f_{\text{low}},f_{\text{high}})$ of
the cluster identified in the time-frequency plane. 
When multiple triggers from one detector are clustered in time
as described in section \ref{sec:clustering} above,
the inclusive frequency band for that clustered event trigger is formed.
For \mycaps{TFCLUSTERS} triggers only,
we require that the frequency bands 
of the clustered event triggers from each detector 
in the triplet either overlap,
or are separated in frequency
space by no more than a fixed window of $\Delta f = 80$ Hz,
based on studies of the simulations described in section~\ref{sec:simulations}.



\section{Background and signal rates}\label{sec:background}

The data processing pipeline (section~\ref{sec:pipeline})
generates background event triggers originating
in noise level fluctuations in the detectors,
due to random processes or environmental or instrumental disturbances.
Our primary means to reject such background event triggers is temporal 
coincidence between the three detectors in the LIGO array
(section \ref{sec:coincidence}).
To the extent that noise fluctuations in each of the detectors
are random, uncorrelated, and follow Poisson statistics,
the primary background comes
from accidental coincident events, and the
accidental triple-coincidence rate can be predicted from the
observed instantaneous single-detector event rates.

\subsection{Background estimation}\label{sec:backestimate}

We have chosen to tune our event trigger generators
(using the playground data sample)
so as to produce an estimated 
accidental triple-coincidence rate of one event over
the entire S1 observation time,
as discussed in section \ref{sec:ETGs}.

Again assuming no correlations between noise fluctuations
in the three detectors, we can indirectly measure the 
rate of accidental triple-coincident events
from triple-coincidence rates when artificial time shifts
are introduced between single-detector event triggers. 

Such time shifted triple-coincidence events are free of
contamination from true gravitational wave bursts
(assuming that such bursts are rare),
and thus are an unbiassed estimate of the 
accidental triple-coincidence rate.
The distribution in the number of 
time shifted triple-coincidence events should follow a
Poisson distribution.
These distributions
can be fitted to obtain
the expected number of background events for use in
our statistical analysis.

The time shifts should be larger than the maximum duration of 
a real (noise-induced or gravitational wave-induced)
detectable burst, or else the events will be correlated
and will not obey Poisson statistics.
The time shifts should also be shorter than the typical time scale
over which the single-detector event rates vary substantially,
so that the number of events for
different time shifts
will be Poisson distributed for a quasi-stationary process.

\begin{figure}[!thb]
\includegraphics[width=0.95\linewidth]{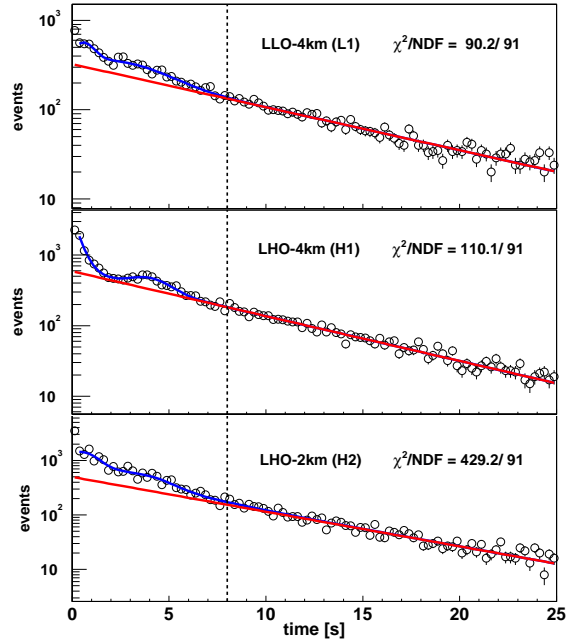}
\caption{Histograms of the time delay between consecutive events
in the \mycaps{TFCLUSTERS} event trigger generation, for the 
L1, H1, and H2 detectors.
The curves are components of fits to the distributions
that incorporate the expectations for short time delay correlations
and long time delay random, uncorrelated events.
The vertical dashed lines indicate the time delay beyond which
consecutive events are consistent with being uncorrelated.
}\label{fig:tfcDel} 
\end{figure}

To establish a lower limit on the time shift required to ensure
uncorrelated noise event triggers, we histogram
the time delay  between consecutive events
in the three detectors, shown in Fig.~\ref{fig:tfcDel}
for the \mycaps{TFCLUSTERS} event trigger generator
(the distributions are similar for the
\mycaps{SLOPE} event trigger generator).
The distributions of delay times follows the expected
exponential form for delay times exceeding 8 seconds
(vertical dashed lines), for all three detectors.
Any residual 
auto-correlations present in the data
will rapidly decay for delay times exceeding 8 seconds,
and in the case of many ($N$) time-shift experiments,
their potentially biased contribution to the Poisson estimate reduce
as $1/N$.

The assumption that noise fluctuations are uncorrelated
between detectors is questionable for the two
detectors co-located at the Hanford site \cite{ref:LHO},
H1 and H2. Indeed, there exists evidence for short-term,
narrow-band correlations in the noise power between the H1 and H2 
detectors associated with power line harmonics,
as well as correlations between L1 and H1 or H2
associated with harmonics of the data acquisition 
buffer rate \cite{ref:SUL}.
The power line harmonics integrate away over long time scales,
and the data acquisition buffer rate harmonics only appear
after long integration times.
It is the short term correlations that concern us here.
We have found no detectable evidence of short term
correlated noise fluctuations associated
with these sources of narrow-band correlations.

In order to account for any potential correlations in noise fluctuations
between H1 and H2, we have performed our time-shifted coincidence
measurements by shifting the time between event triggers
found in the L1 data and those found in the H1 and H2 data,
while keeping zero time shift between H1 and H2.

\begin{figure}[!thb]
\includegraphics[width=0.95\linewidth]{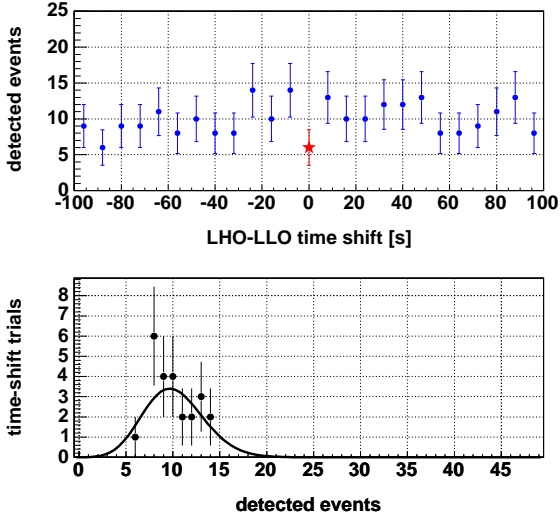}
\caption{Time-shifted triple coincident events from
\mycaps{TFCLUSTERS} event triggers, as a function of 
an artificial time shift introduced
between the Hanford (LHO) and Livingston (LLO) sites.
Top: Number of events versus time shift,
in 8 second steps; the point at zero time shift 
is the number of true triple coincident events.
Bottom: Histogram of the number of time-shifted coincident events,
with the Poisson fit overlaid (the zero time shift point is excluded).
In both plots, the error bars are Poissonian.
}\label{fig:tfcLag} 
\end{figure}

\begin{figure}[!thb]
\includegraphics[width=0.95\linewidth]{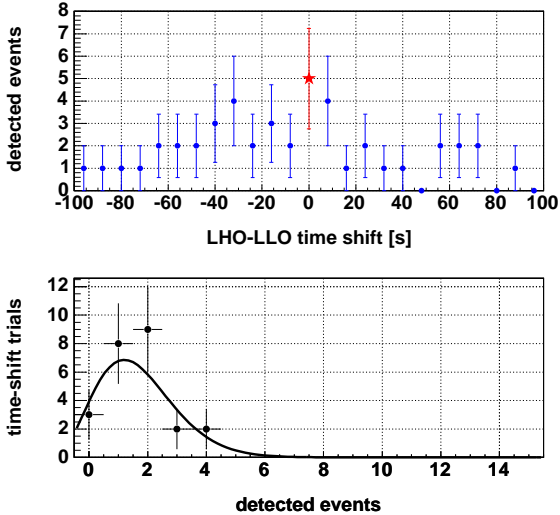}
\caption{Same as Fig.~\ref{fig:tfcLag},
but for the \mycaps{SLOPE} event trigger generator.
}\label{fig:slopeLag} 
\end{figure}

We have performed multiple time-shift experiments with the \mycaps{SLOPE} and
\mycaps{TFCLUSTERS} event trigger generators
between the Livingston and Hanford sites.
The resulting number of time-shifted triple coincident events
from 24
such experiments in the $[-100,100]$ second range with 8
second steps are shown in Figs.~\ref{fig:tfcLag} and \ref{fig:slopeLag}
for \mycaps{TFCLUSTERS} and \mycaps{SLOPE}, respectively.
The distributions of background events for the 24 non-zero time shifts
(lower plots in Figs.~\ref{fig:tfcLag} and \ref{fig:slopeLag})
are fitted with Poisson predictions and are found to be
consistent with the expectation from Poisson statistics.
Averages and Poisson mean values for different
step and window sizes vary by less than 0.5 events.

In estimating the background rate,
we have considered time shifts between 8 seconds
(to avoid correlated events; see Fig.~\ref{fig:tfcDel}) 
and 100 seconds
(to minimize dependence on any
non-stationarity in the instantaneous event rate).
These time-shift measurements yield estimates
of the number of
accidental triple-coincidence (background) events $\mu_B$ 
for the \mycaps{TFCLUSTERS} and \mycaps{SLOPE}
event triggers.
Because these measurements can be made with many,
statistically-independent time shifts,
the resulting statistical errors on these estimates
are small. 

The presence of any remaining non-stationarity in the 
background event rate, however, will result in 
errors in the background rate estimation.
In fact, the instantaneous event trigger rate is observed
to vary for both event trigger generators.
The variability of the trigger rate is sufficiently
severe for \mycaps{SLOPE} that the background rate estimation
is deemed unreliable.
Because of the fixed (non-adaptive)
threshold employed in the \mycaps{SLOPE} algorithm,
the trigger rate of the individual
interferometers varies by more than a factor of one thousand, 
sometimes on timescales of 10 seconds or less.
All five events of the zero time lag coincidences
and most of the time-shifted coincidences in Fig.~\ref{fig:slopeLag}
come from
a single 360-second segment, corresponding to a
coincidence of data segments from three interferometers with
very high burst singles rates. Even within
that segment, the singles rate varies markedly, 
making it difficult to reliably estimate the background rate.
For this reason, we choose not to use the 
\mycaps{SLOPE} pipeline to set a limit on
the rate of gravitational wave bursts~\cite{ref:meaculpa}.

It should be noted that
before these coincidences and background rates were available, 
we decided to set our upper limits using the results from the
event trigger generator which yielded
the better efficiency for detecting gravitational wave bursts, 
as measured by our simulations.  
For almost all waveforms this turned out to be
the \mycaps{TFCLUSTERS} pipeline (see section \ref{sec:sensitivity}).
Thus, even if the background rate (and thus the rate of 
excess triggers) from the \mycaps{SLOPE} pipeline
could be reliably estimated, the primary results from
this search would still be based on the \mycaps{TFCLUSTERS} pipeline.

\subsection{Signal candidate estimation}\label{sec:foreground}

An excess in the number of coincident (zero-time-shift) events 
over the estimated background can be estimated statistically.
Here we make use of the unified approach
of Feldman and Cousins~\cite{feldman98a}.
This approach provides an unambiguous prescription for 
establishing a statistical excess of signal candidate events
at a specified confidence level (that is, a lower limit 
to the confidence interval that is greater than zero).
However, as discussed in section~\ref{sec:intro},
we have not yet characterized our detectors and data analysis procedures
sufficiently well to claim that any such excess is a 
detection of gravitational wave bursts.
We therefore use only the upper endpoint on the
confidence interval for the number of 
signal candidate events to set an
upper limit on the rate of gravitational wave bursts. 

Starting from an observed number of events $n$
and an estimate of the number of background events $\mu_B$,
we build confidence bands 
for the number of signal events $\mu_S$
according to the formula:
\begin{equation}
\sum_{n_0=n_1}^{n_2} p(n_0)_{\mu_S+\mu_B} \ge \alpha,
\label{eq:FC}
\end{equation}
where $p(n)_{\mu_S+\mu_B}$
is the Poisson probability density function
\begin{equation}
p(n)_{\mu_S+\mu_B}=(\mu_S+\mu_B)^n\frac{e^{-(\mu_S+\mu_B)}}{n!}.
\end{equation}
The sum extremes, $n_1$ and $n_2$,
are chosen according to a likelihood ranking principle~\cite{feldman98a}.
In our implementation, we assume both signal and background are
Poisson distributed.
We report confidence bands for
$\alpha = 90\%$, $95\%$, and $99\%$.

We account for the statistical error on the background
estimation following the method described in~\cite{conrad03a,conrad02a},
where a Gaussian background uncertainty is folded in the formulation of the 
probability density function. We replace $p(n)_{\mu_S+\mu_B}$
in Eqn.~\ref{eq:FC} with
\begin{equation}
  q(n)_{\mu_S+\mu_B}=\frac{1}{\sqrt{2\pi}\sigma_B}
  \int_0^\infty
  p(n)_{\mu_S+\mu_B'}\,e^{-\frac{(\mu_B-\mu_B')^2}{2\sigma_B^2}}\,
  d\mu_B' 
\end{equation}
where
$\sigma_B$ is the estimated background error.
This marginalization, performed through a Monte Carlo calculation, is
used in the construction of confidence bands for the estimated
background $\mu_B\pm\sigma_B$.

\subsection{Event rate bound}\label{sec:bound}

Table \ref{tbl:S1raw} shows the number of coincident events,
the estimated number of accidental coincident events (background),
and the confidence bands that we find at the
90\%, 95\%, and 99\%\ confidence levels
on the number of excess events and the event rate (over 35.5 hours
of S1 observation time), using
the \mycaps{TFCLUSTERS} event trigger generator pipeline.
The upper bounds of the confidence bands are taken to be the upper limit
on the number of signal events, at that confidence level.
At the 90\%\ confidence level,
the search yields an upper limit of 2.3
events in 35.5 hours.
As discussed in section~\ref{sec:backestimate},
because of the variability of the event trigger rate 
in the \mycaps{SLOPE} pipeline,
we choose not to use it to set a limit on the rate
of gravitational wave bursts.

\begin{table}[!htb]
\caption{Confidence bands
on the number of excess events in the S1 run (35.5 hours of observation time)
from the \mycaps{TFCLUSTERS} pipeline.}
\label{tbl:S1raw}
\begin{tabular}{lcc}
\hline\hline
Coincident events      &              6 \\
Background             & $10.1 \pm 0.6$ \\
90\% confidence band   &      $0 - 2.3$ \\
95\% confidence band   &      $0 - 3.5$ \\
99\% confidence band   &      $0 - 5.9$ \\
\hline\hline
\end{tabular}
\end{table}

Given the estimated
backgrounds from the time shift analyses, 
the number of \mycaps{TFCLUSTERS} events
at zero time lag is somewhat low 
(Table~\ref{tbl:S1raw} and Fig.~\ref{fig:tfcLag}).
None of the events detected by \mycaps{SLOPE} were detected
by \mycaps{TFCLUSTERS}.  This is not in itself surprising, since the two 
event trigger generators have
different sensitivities to different waveforms, but it does indicate that
none of the events were far above threshold for that 
trigger generator, since the largest
differences in efficiency between the two event trigger generators
was approximately a factor of 6 (section~\ref{sec:sensitivity} below).
The probability of obtaining six or fewer \mycaps{TFCLUSTERS} events, 
given our estimated background, is approximately 12\%.   
We found no reason to suspect any systematic errors in our background estimate
for this pipeline.  Alternative methods of estimating the background
(simple estimates based on the average singles rates and the coincidence
window, time shift analyses where all three detectors are shifted as
opposed to holding H1-H2 fixed at zero delay) did not give significantly
different background rates.



\section{Efficiency determination}\label{sec:efficiency}

In order to interpret our bound on the observed rate 
for coincident gravitational wave bursts,
we study the response of the LIGO detectors and 
our analysis pipeline to simulated signals 
with varying waveforms, durations, bandwidth, and peak amplitudes.
The simulated signals were injected into the gravitational wave
data stream from each of the three detectors,
as far upstream in the pipeline as was practical
(after data acquisition and ingestion into the
LIGO Data Analysis System).

The same data that were used to search for coincident bursts
(section \ref{sec:preselection}) were also used for these simulations;
and, for the purposes of these simulations,
these data are assumed to consist entirely of noise
(no real gravitational wave bursts present).
Approximately 20\%\ of the S1 data was used
for these simulations, spanning the entire data run uniformly,
in order to fairly represent the noise and 
detector sensitivities throughout the run.

We present the results for the efficiency determinations for
both \mycaps{TFCLUSTERS} and \mycaps{SLOPE}
event trigger generators, even though 
(as noted in section~\ref{sec:backestimate})
we do not use \mycaps{SLOPE}
to derive a final limit on the rate of gravitational
wave bursts.

\subsection{Waveforms}\label{sec:waveforms}

The astrophysical origin, and waveform morphology,
of the gravitational wave bursts we search for in this work
are a priori unknown.
A broad range of signal waveforms were considered.
These include astrophysically-motivated waveforms,
such as the results of supernova simulations
\cite{ref:ZM,ref:DFM},
as well as ad-hoc waveforms such as Gaussians,
damped sinusoids, sine-Gaussians, Hermite-Gaussians,
and others.
Guided by the simulations in \cite{ref:ZM,ref:DFM},
we have endeavored to be
sensitive to any waveform that adds excess power
(over that of the detector noise) in the LIGO S1
sensitivity band (150 to 3000 Hz), 
with durations between 4 ms and 100 ms.

\begin{figure}[!thb]
\includegraphics[width=0.95\linewidth]{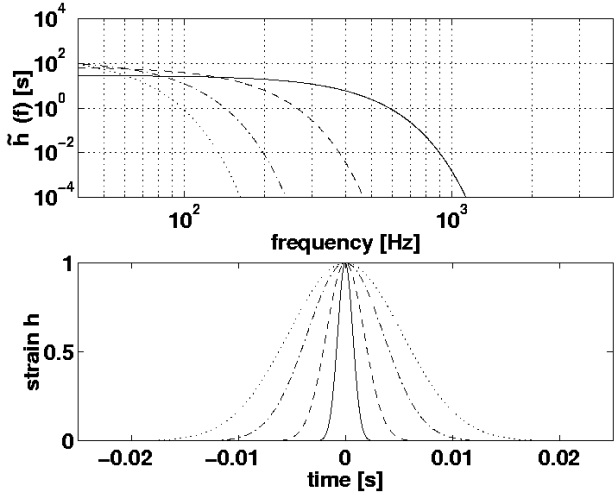}
\caption{Gaussian waveforms, with varying duration,
as described in section~\ref{sec:waveforms}.
Top: Frequency spectrum. Bottom: Time series.
}\label{fig:Gaussian}
\end{figure}

\begin{figure}[!thb]
\includegraphics[width=0.95\linewidth]{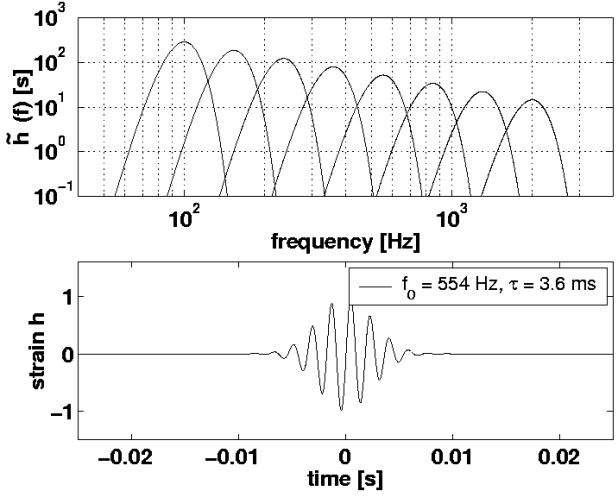}
\caption{Sine-Gaussian waveforms, 
with central frequency f$_0$ evenly spaced in log(f),
and $Q \sim 9$,
as described in section~\ref{sec:waveforms}.
Top: Frequency spectrum. Bottom: Time series for 
a sine-Gaussian with f$_0$ = 554 Hz.
}\label{fig:SineGauss}
\end{figure}

In order to evaluate our sensitivity to such bursts,
we must model the waveforms in some general way.
For the results presented here, we have chosen
to focus on two classes of ad hoc waveforms,
which we regard as ``surrogates'' for real astrophysical signals.
The first are broad-band, limited-duration 
Gaussians of the form
\begin{equation}
h(t+t_0) = h_0 \exp\left(-t^2/\tau^2\right),
\label{eq:GA}
\end{equation}
with varying peak amplitude $h_0$, peak time $t_0$,
and duration $\tau$ (Fig.~\ref{fig:Gaussian}).
The second are narrower-band,  limited-duration
sine-Gaussians of the form
\begin{equation}
h(t+t_0) = h_0 \sin\left(2\pi f_0 t\right)\exp\left(-t^2/\tau^2\right).
\label{eq:SG}
\end{equation}
The duration of the sine-Gaussians were chosen
to be $\tau = 2/f_0$. Their Fourier transforms $\tilde{h}(f)$
span a (Gaussian) frequency band of $\sigma_f = f_0/Q$
centered about the central frequency $f_0$,
where $Q \equiv f_0/\sigma_f =\sqrt{2}\pi \tau f_0 \simeq 8.9$.
We have chosen eight different central frequencies,
spaced logarithmically, and spanning the LIGO 
sensitivity band: 
$f_0 = [100, 153, 235, 361, 554, 850, 1304, 2000]$ Hz
(Fig.~\ref{fig:SineGauss}).

Our analysis pipeline detection efficiency 
depends on the burst duration, frequency band,
and some measure of burst ``strength'';
it does not depend strongly on the precise 
waveform morphology.
In order to facilitate comparison of the burst strength
with the detectors' equivalent strain noise,
and with burst waveforms having similar properties but
different detailed morphologies,
we define two useful measures of the burst strength.
The {\it root-sum-square ({\it rss})}
amplitude spectral density for 
such bursts, in units of dimensionless strain per root Hz, is defined by:
\begin{eqnarray}
 h_{rss} &\equiv & \sqrt{\int |h|^2 dt}; \label{eq:hrss}\\
         & =     & \sqrt{\sqrt{\pi/2} \tau} h_0 \,\mbox{(Gaussians)}; \label{eq:hrssGA}\\
         & =     & \sqrt{Q/(4\sqrt{\pi} f_0)} h_0 \,\mbox{(sine-Gaussians)}.\label{eq:hrssSG}
\end{eqnarray}
The {\it characteristic} strain
amplitude, in units of dimensionless strain, 
is defined by~\cite{ref:FlanHughes}:
\begin{eqnarray}
 h_{char} &\equiv & f_c |\tilde{h}(f_c)| ; \label{eq:hc} \\
         & =     & \sqrt{\pi} (f_c\tau) \exp\left(-(\pi f_c\tau)^2\right) h_0  \,\mbox{(Gaussians)}; \label{eq:hcGA}\\
         & =     & \sqrt{\pi} (f_c\tau /2) h_0 \,(\mbox{sine-Gaussians}, f_c = f_0).\label{eq:hcSG}
\end{eqnarray}
Here, $\tilde{h}(f)$ is the Fourier transform of $h(t)$, defined by
\begin{equation}\label{eq:fourier}
 \tilde{h}(f) = \int^\infty_{-\infty} h(t) e^{-i2\pi ft} dt,
\end{equation}
and $f_c$ is a characteristic frequency (typically, 
either the frequency at which $\tilde{h}(f)$ peaks,
or the frequency where the detector is most sensitive).
For our sine-Gaussians, we choose $f_c$ to be 
the central frequency $f_0 = 2/\tau$; for Gaussians, we choose 
$f_c$ to be the frequency at which all three LIGO detectors
had approximately best sensitivity during S1, $f_c \simeq 300$ Hz
(see Fig.~\ref{fig:strain}).

\subsection{Simulations}\label{sec:simulations}

In order to add the simulated signal (in units of 
dimensionless strain) to the raw detector data
(in units of ADC counts), we must convert,
or filter, the signal using the detector response function
(in counts per strain) obtained
through the calibration procedure described 
in section \ref{sec:detectors}.
The simulated signals, padded with zeros to minimize
edge effects, are filtered through the detector 
response function in the Fourier domain,
yielding a time series in ADC counts
that can be added directly to the raw gravitational wave
data stream at the beginning of the data processing pipeline.
These simulated signals can be injected
at any chosen point in time, and with any chosen amplitude.
The uncertainty in the calibration information
is the largest source of systematic error in this analysis
(section~\ref{sec:systematics}).

\begin{figure}[!thb]
\includegraphics[width=0.95\linewidth]{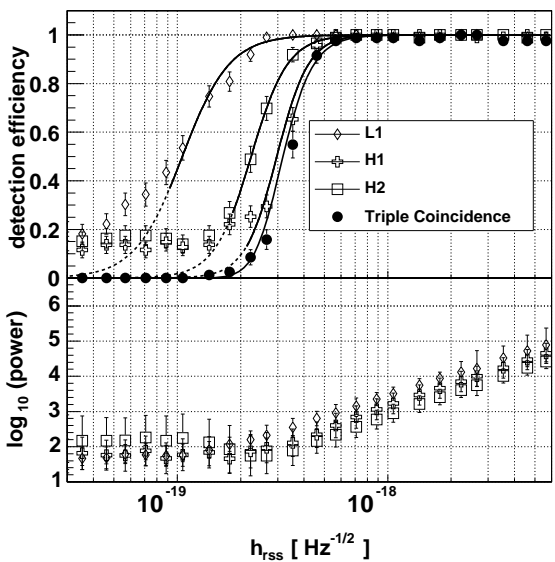}
\caption{
The response of the \mycaps{TFCLUSTERS} event trigger generator to
Gaussian bursts with $\tau = 1$ ms, 
embedded in S1 data,
as a function of the root-sum-square strain $h_{rss}$.
Upper plot: Average burst detection efficiency.
The efficiencies were evaluated through simulations
of burst waveforms with optimal wave direction and polarization,
injected into S1 data.
The simulated data points are fitted to sigmoid curves, shown,
in the region where the efficiency is not dominated
by random noise triggers.
The curve for the triple-coincidence is the product of the single-detector 
efficiency curves, and can be directly compared with the
triple-coincidence simulation data points.
Lower plot: Average
detected signal strength for each of the three LIGO detectors.
}\label{fig:eff1}
\end{figure}

\begin{figure}[!thb]
\includegraphics[width=0.95\linewidth]{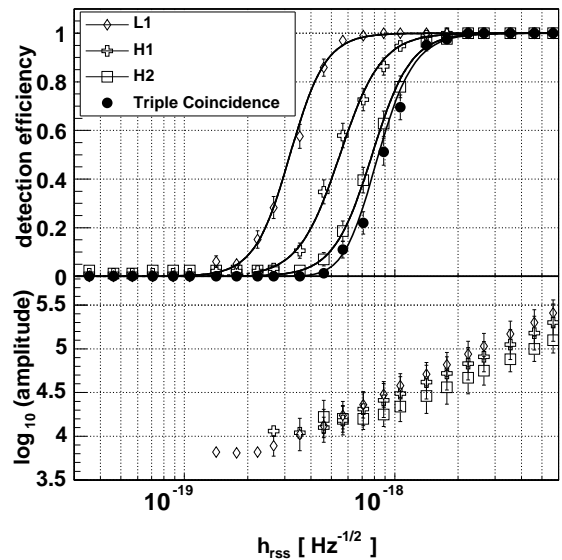}
\caption{This is the same as Fig.~\ref{fig:eff1} 
but obtained with the \mycaps{SLOPE} event trigger generator.
}\label{fig:eff1slope}
\end{figure}

For each waveform, we evaluate the efficiency for 
detection through each of the three LIGO detectors
and analysis pipelines, as a function of 
$h_{rss}$ (defined in equation~\ref{eq:hrss}),      
assuming optimal wave direction and polarization.
Approximately 80 simulations are performed 
for each combination of waveform, $h_{rss}$,        
detector, and event trigger generator,
using data spanning the S1 run.
In Figs.~\ref{fig:eff1} and ~\ref{fig:eff1slope}
we plot detection efficiencies and average signal strengths 
for the \mycaps{TFCLUSTERS} and \mycaps{SLOPE} 
event trigger generators, respectively.
Although our event trigger generators
do not necessarily trigger on excess power,
we find that the ``strength'' of the signal
reported by either event trigger generator
(the  \mycaps{SLOPE} ``amplitude'' or the
square root of the \mycaps{TFCLUSTERS} ``power'')
is proportional to the actual amplitude of the injected signal
over a broad range of $h_{rss}$.                   
This is illustrated by the lower plots in
Figs.~\ref{fig:eff1} and ~\ref{fig:eff1slope}
(for one particular waveform).
We emphasize that the ``power'' or ``amplitude'' that is plotted
in Figs.~\ref{fig:eff1} and ~\ref{fig:eff1slope}, respectively
(and in Figs.~\ref{fig:tfcHist} and \ref{fig:slopeHist}),
are purely algorithm-dependent quantities which are compared
with thresholds to define event triggers;
they are not designed to be true measures of the
burst power or amplitude,
and they will be different for different detectors and 
waveforms.

As expected, the efficiencies are essentially 100\%\ for 
large values of $h_{rss}$,                          
consistent with noise and thus
0\%\ efficiency for small $h_{rss}$,               
and transitioning smoothly
over a narrow intermediate range of $h_{rss}$.      
The time window used 
to associate a \mycaps{TFCLUSTERS} event trigger around the
time of the injection (0.5 s) is larger than for 
a \mycaps{SLOPE} event trigger, so
the observed efficiency for 
small $h_{rss}$ waveforms appears larger in Fig.~\ref{fig:eff1}
than in Fig.~\ref{fig:eff1slope}.
The results in both cases are empirically found to be well
fitted to simple sigmoid curves in
$\log_{10}(h_{rss})$:                               
\begin{equation}\label{e:sigmoid}
\varepsilon(h_{rss})=\frac{1}{1+\mbox{e}^{-(\log_{10}{h_{rss}}-b)/a}}
\end{equation}
where $b = \log_{10}{h_{rss\, 1/2}}$                
determines the strain per root Hz                   
$h_{rss\, 1/2}$
at which the efficiency is equal to 1/2, 
and $a$ governs the width of the transition
from 0 to 1 in $\log_{10}(h_{rss})$.                
It is specific to a given waveform, 
detector, event trigger generator, and data epoch.
All fits resulted in good fit quality,
except at the smallest values of $h_{rss}$,         
where noise triggers dominate;
we exclude such triggers from our definition of ``efficiency'',
and use the sigmoid fits to extrapolate to 
zero efficiency at small values of $h_{rss}$.
Examples of sigmoid fits are shown in Fig.~\ref{fig:eff1}
and Fig.~\ref{fig:eff1slope}.

\def\rapprox{\sim\kern-1em\raise 0.6ex\hbox{$>$}}

The efficiencies shown in 
Figs.~\ref{fig:eff1} and \ref{fig:eff1slope}
do not remain
at unity to arbitrarily high amplitudes.  A sufficiently strong
gravitational wave could trigger the BLRMS cut and in that way
effectively prevent its own detection in this search.  The most
susceptible band for such a possibility is the 320-400 Hz band
in L1, where, for example, a 361 Hz  sine-Gaussian with
$h_{rss} \rapprox 6 \times 10^{-18}$ could begin to
trigger the BLRMS cut.   Signals centered at other frequencies
or those with broader bandwidths would require a higher amplitude.
We estimate that a loud supernova \cite{ref:ZM,ref:DFM}
at 7 pc, or a 3+3 solar mass
binary neutron star inspiral at 300 pc could begin to trigger the
BLRMS cut. (Note however that the well-defined waveform of the
latter makes a template-based search \cite{ref:IUL}
a more sensitive method
for detecting such waves.)  The need for this cut in the data was
driven by the nonstationarity of the noise in the detectors during S2,
and detector improvements are expected to reduce our use of
such cuts in the future.

In Fig.~\ref{fig:strain}
we compare the value of $h_{rss}$ for which our simulations
of sine-Gaussian waveforms (at optimal wave direction and polarization)
yield 50\%\ efficiency (averaged over the entire S1 run),
shown as circles,
with the detectors' (typical) equivalent strain noise.
These 50\%\ efficiency points are roughly an order of magnitude
larger than the equivalent strain noise.

\subsection{Average over direction and polarization}\label{sec:direction}

The response of a LIGO detector to an incident 
gravitational wave burst depends on the 
wave direction and wave polarization
relative to the detector axes, and is
referred to as the detector's antenna pattern \cite{thorne87a}.
The only effects of 
the wave direction and polarization are to modify
the amplitude of the detected wave
and the relative arrival times at the detectors.
Since we have evaluated the detection efficiency
for each detector as a function of the root-sum-square strain 
of the wave at optimal direction (directly overhead)
and polarization (aligned with the detector axes),
it is straightforward to evaluate the efficiency 
at arbitrary direction and polarization.
We choose to consider a population of sources
distributed isotropically in the sky,
with random linear polarization.
We thus evaluate the detection efficiency
averaged over direction and polarization,
as a function of intrinsic strain per root Hz  
incident on the Earth.

\begin{figure}[!thb]
\includegraphics[width=0.95\linewidth]{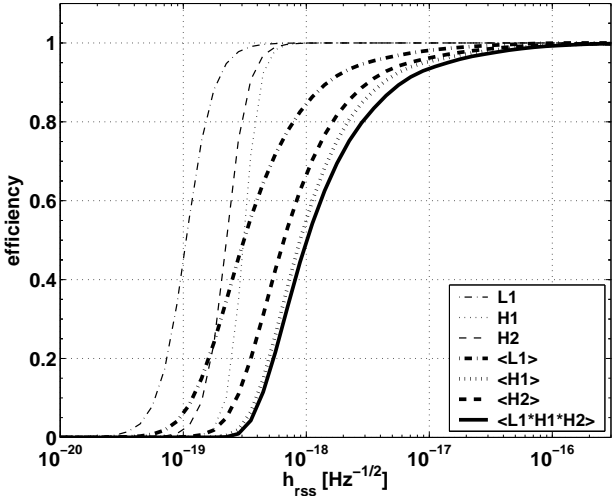}
\caption{Burst detection efficiency for Gaussian bursts
with $\tau = 1$ ms, as a function of 
$h_{rss}$, for each of the three LIGO detectors,
and for the triple coincidence, 
using the \mycaps{TFCLUSTERS} event trigger generator.
The lighter grey curves are the same as the curves in Fig.~\ref{fig:eff1}.
The darker curves to the right of them
are the result of averaging 
the efficiency curves over wave directions and polarizations
(denoted by $\left<\cdots\right>$ in the legend)
as described in section~\ref{sec:direction}.
\label{fig:effavg}}
\end{figure}

\begin{figure}[!thb]
\includegraphics[width=0.95\linewidth]{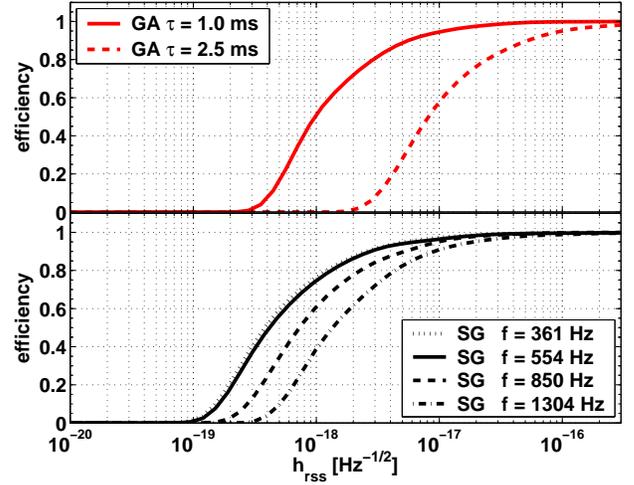}
\caption{Burst detection efficiency for 
triple coincidence as a function of $h_{rss}$, 
using the \mycaps{TFCLUSTERS} event trigger generator,
averaging over wave directions and polarizations,
for six different waveforms: GA refers to the Gaussians defined in
Eqn.~\ref{eq:GA} and SG to the sine-Gaussians defined in
Eqn.~\ref{eq:SG}.
}\label{fig:effwavesTF}
\end{figure}

\begin{figure}[!thb]
\includegraphics[width=0.95\linewidth]{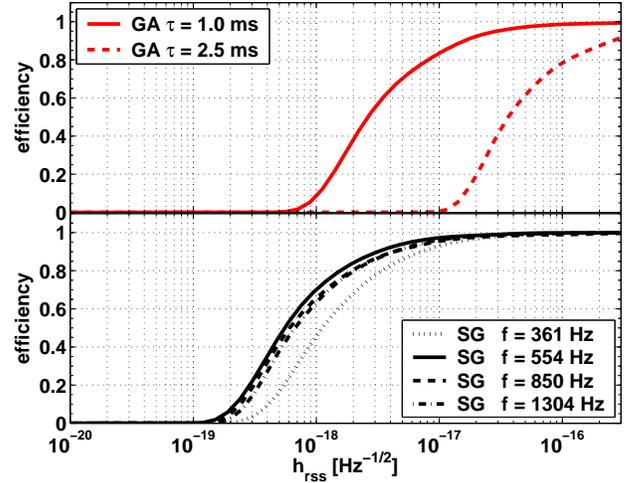}
\caption{\protect Same as Fig.~\ref{fig:effwavesTF},
for the \mycaps{SLOPE} event trigger generator.}\label{fig:effwavesSL}
\end{figure}

In order to evaluate the efficiency for coincident
detection by non-colocated detectors,
we assume that the detection efficiency is 
a measure of a random process, uncorrelated between
detectors. Further, the difference in arrival times
at the different detectors is small compared to the
time coincidence window employed (section~\ref{sec:temporalCoinc}).
Therefore, the efficiency for triple coincidence
can be expressed as the product of efficiencies
for the three LIGO detectors evaluated 
at the appropriate peak amplitude for each.
Under this assumption, the efficiency for coincident
detection by all three LIGO detectors,
averaged over wave direction and polarization,
can be evaluated numerically. 
The results of this procedure are shown in
Figs.~\ref{fig:effavg},
\ref{fig:effwavesTF}, and \ref{fig:effwavesSL}.

The single-detector efficiencies will be
independent only if there are no significant
noise correlations between the detectors.
We have compared the results for coincident detection
to direct simulations
of triple coincidence through the full three-detector
analysis pipeline 
(under the simplifying assumption of neglecting the difference
in antenna pattern response between the two sites),
and found good agreement
(see triple coincidence data points in Figs.~\ref{fig:eff1}
and \ref{fig:eff1slope});
no evidence for burst detection efficiency
correlations between the detectors has been observed.

A crucial test of the accuracy of this simulation procedure
is the comparison of signals injected into the data stream 
with software,
with signals injected directly into the end test masses
of the interferometers.
The comparison of these ``hardware'' burst injections with 
the (much more numerous) software injections
provide a test of the detector response,
the calibration information, the data acquisition,
and the entire software analysis chain,
including the software simulations used to evaluate
the efficiency, as described above.

Hardware injections
of sine-Gaussians with varying $f_0$ and $h_{rss}$           
were performed during brief periods, just prior to 
the beginning and just after the end of the S1 run.
Due to time constraints, only a limited number of 
hardware injections were performed.
As a result, the comparison with software injections
is somewhat qualitative.
The detection of these hardware
injections through the analysis pipeline was found to 
be consistent with expectations from
the software injections.

\subsection{Sensitivity to modeled bursts}\label{sec:sensitivity}

We can use the efficiency determined by simulated signal injections,
discussed in sections~\ref{sec:simulations} and \ref{sec:direction},
to estimate the weakest signal we could have seen in the 
search described in this article.
The efficiencies of each of our two event trigger
generators for several different waveforms
is shown in Figure \ref{fig:effwavesTF} and 
Figure \ref{fig:effwavesSL}. 
The sensitivity at 50\%\ efficiency,
for a variety of Gaussian and sine-Gaussian waveforms,
is shown in Table~\ref{tbl:S1sens}
in terms of $h_{rss}$,
and in Table~\ref{tbl:S1sensc}
in terms of $h_{char}$.

\begin{table}[!htb]
\caption{
Sensitivity to various waveforms in the S1 run from 
\mycaps{TFCLUSTERS} and \mycaps{SLOPE} pipelines
for triple-coincidence detection,
averaged over source direction and polarization.
The sensitivity is given in terms of $h_{rss}$
(Eqn.~\ref{eq:hrss}, units of Hz$^{-1/2}$), at 50\%\ efficiency
($h_{rss\, 1/2}$).
A 20\%\ uncertainty associated with calibration
(section \ref{sec:systematics}) is not included.
}
\label{tbl:S1sens}
\begin{tabular}{lcc}
\hline\hline
 & \mycaps{TFCLUSTERS} & \mycaps{SLOPE} \\
 & $[ {\rm Hz}^{-1/2}]$ & $[ {\rm Hz}^{-1/2}]$  \\
\hline
Gaussian $\tau = 1.0\,$ms      & $1.0\times 10^{-18}$ & $2.6\times 10^{-18}$ \\
Gaussian $\tau = 2.5\,$ms      & $8.2\times 10^{-18}$ & $3.6\times 10^{-17}$ \\
sine-Gaussian $f_0 = 153\,$Hz  & $1.6\times 10^{-18}$ & $1.2\times 10^{-17}$ \\
sine-Gaussian $f_0 = 235\,$Hz  & $5.1\times 10^{-19}$ & $2.8\times 10^{-18}$ \\
sine-Gaussian $f_0 = 361\,$Hz  & $3.8\times 10^{-19}$ & $1.1\times 10^{-18}$ \\
sine-Gaussian $f_0 = 554\,$Hz  & $4.2\times 10^{-19}$ & $5.6\times 10^{-19}$ \\
sine-Gaussian $f_0 = 850\,$Hz  & $7.3\times 10^{-19}$ & $6.1\times 10^{-19}$ \\
sine-Gaussian $f_0 = 1304\,$Hz & $1.4\times 10^{-18}$ & $6.7\times 10^{-19}$ \\
sine-Gaussian $f_0 = 2000\,$Hz & $2.3\times 10^{-18}$ & $2.5\times 10^{-18}$ \\
\hline\hline
\end{tabular}
\end{table}

\begin{table}[!htb]
\caption{
Sensitivity to various waveforms in the S1 run from 
\mycaps{TFCLUSTERS} and \mycaps{SLOPE} pipelines
for triple-coincidence detection,
averaged over source direction and polarization.
The sensitivity is given in terms of $h_{char}$
(Eqn.~\ref{eq:hc}, dimensionless strain), at 50\%\ efficiency.
A 20\%\ uncertainty associated with calibration
(section \ref{sec:systematics}) is not included.
}
\label{tbl:S1sensc}
\begin{tabular}{lcc}
\hline\hline
 & \mycaps{TFCLUSTERS} & \mycaps{SLOPE} \\
\hline
Gaussian $\tau = 1.0\,$ms      & $1.4\times 10^{-18}$ & $3.6\times 10^{-18}$ \\
Gaussian $\tau = 2.5\,$ms      & $3.3\times 10^{-19}$ & $1.5\times 10^{-18}$ \\
sine-Gaussian $f_0 = 153\,$Hz  & $3.1\times 10^{-17}$ & $2.4\times 10^{-16}$ \\
sine-Gaussian $f_0 = 235\,$Hz  & $1.2\times 10^{-17}$ & $6.8\times 10^{-17}$ \\
sine-Gaussian $f_0 = 361\,$Hz  & $1.1\times 10^{-17}$ & $3.3\times 10^{-17}$ \\
sine-Gaussian $f_0 = 554\,$Hz  & $1.6\times 10^{-17}$ & $2.1\times 10^{-17}$ \\
sine-Gaussian $f_0 = 850\,$Hz  & $3.4\times 10^{-17}$ & $2.8\times 10^{-17}$ \\
sine-Gaussian $f_0 = 1304\,$Hz & $8.0\times 10^{-17}$ & $3.8\times 10^{-17}$ \\
sine-Gaussian $f_0 = 2000\,$Hz & $1.6\times 10^{-16}$ & $1.8\times 10^{-16}$ \\
\hline\hline
\end{tabular}
\end{table}



\section{Interpreted Results}\label{sec:results}

\subsection{Exclusion in rate versus strength plane}

\begin{figure}[!thb]
\includegraphics[width=0.95\linewidth]{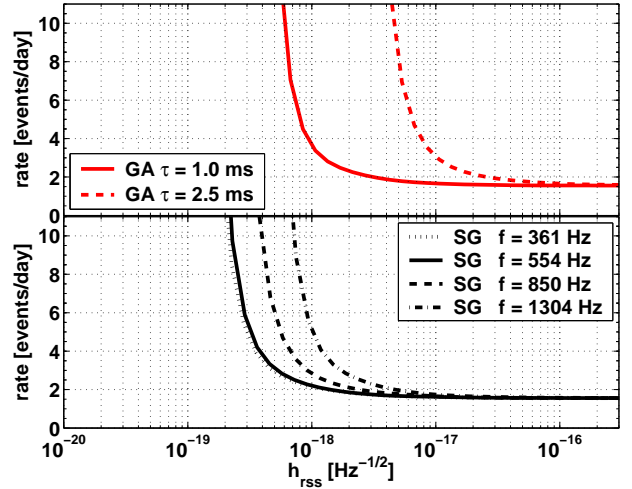}
\caption{Rate versus $h_{rss}$ for detection
of specific waveforms
using the \mycaps{TFCLUSTERS} event trigger generator.
The region above and to the right of the curves
is excluded at 90\%\ confidence level or greater.
The effect of the 20\%\ uncertainty in the detector response
is included.
Top: For Gaussians with $\tau = 1.0$~ms and $\tau = 2.5$~ms.
Bottom: For sine-Gaussians with $Q=9$ and central frequency
$f_0 =$ 361, 554, 850 and 1304 Hz.
}\label{fig:TFfinal}
\end{figure}

The results of our search can be used to set limits on models of
ensembles of gravitational waves arriving at the earth. 
Fig.~\ref{fig:TFfinal} shows
the upper limits that we set, 
using the \mycaps{TFCLUSTERS} 
event trigger generator,   
as expressed in the plane of event rate
versus $h_{rss}$. 
The top figure is
for the case of 1 ms and 2.5 ms Gaussian bursts,
and the lower figure is for sine-Gaussian bursts with
central frequency of 361, 554, 850 and 1304  Hz.

As discussed in sections \ref{sec:intro} and \ref{sec:efficiency},
these limits are given in terms of
an ensemble of waves of equal amplitude,
incident on the earth from all directions and with 
all (linear) polarizations.
This ensemble is not motivated by astrophysical considerations,
but is nevertheless useful in
characterizing the performance of the search,
and it can be compared with similar limits obtained
by resonant bar detector collaborations \cite{ref:bars2000,ref:bars2001}.

The curves in Fig.~\ref{fig:TFfinal}
are constructed
by dividing our observed event rate upper limit
by the efficiency curves shown in Fig.~\ref{fig:effwavesTF}.
In the limit of large wave amplitudes $h_{rss}$ where our
efficiency is essentially unity, the upper limit is independent of
amplitude, at a level given by the burst rate limit presented
in section~\ref{sec:bound}.
The limit at all amplitudes with lower efficiency is given by
that burst rate limit, multiplied by the inverse of the efficiency.

The meaning of the upper limit curve can be understood by imagining
that one is interested in the flux of 1 ms Gaussian gravitational
wave bursts at a particular amplitude. Reading the value of the curve
at that amplitude gives the 90\%\ confidence level upper limit
on the rate of such bursts with $h_{rss}$ equal to or greater than
that amplitude.
Similarly, one can use these curves to determine
the 90\%\ confidence level upper limit on the 
amplitude of bursts which are incident at a specified rate.

\subsection{Comparing results from the two pipelines}

As discussed in section~\ref{sec:backestimate}, the variability in the 
\mycaps{SLOPE} background
led us to choose not to use it to set an upper limit on the gravitational
wave burst rate.  
However, the two event trigger generators use very different
and complementary methods
to identify bursts in the data, and 
it is thus interesting to compare the results from 
the two pipelines.
We want to make the strongest statement that we can about the rate
of gravitational wave bursts, consistent with sound statistical
practice. 
We evaluated the efficiency of our two 
event trigger generator pipelines 
for each of our candidate signal waveforms,
tuned to give approximately the same background rate 
using the playground data set.
To combine the results from the
two pipelines, we would
choose to quote upper limits from the event trigger generator
that has the better efficiency for the largest number of waveforms.
With the tuning parameters used in this study,
we find that the \mycaps{TFCLUSTERS} pipeline 
has better efficiency than \mycaps{SLOPE} for
most of these waveforms (Table~\ref{tbl:S1sens}),
although \mycaps{SLOPE} performs somewhat better for the 
850 Hz and 1304 Hz sine-Gaussians.
Thus independent of the decision to not use the SLOPE result
because of the problems with background variability,
the final upper limits that we would set are the ones 
obtained from the \mycaps{TFCLUSTERS} pipeline, 
shown in Fig.~\ref{fig:TFfinal}.
The \mycaps{SLOPE} results provide a valuable 
cross-check, and we intend to continue to employ
and improve the \mycaps{SLOPE} pipeline
in future analyses (section~\ref{sec:future}).




\subsection{Systematic uncertainties}\label{sec:systematics}

The following systematic uncertainties are known to significantly
contribute to systematic errors associated with the results of our
search.
The estimation of our efficiency for detection
of bursts with modeled waveforms (the Gaussians
and sine-Gaussians that were  considered in 
section~\ref{sec:simulations})
has a statistical error associated with the finite
number of simulations.
The underlying noise floor was modeled using a 
sample of the S1 data itself; there is a systematic
uncertainty associated with the degree to which this sample
was representative of the entire S1 dataset.

The analysis procedure for the efficiency
has various potential sources of error.
The detection efficiency versus $h_{rss}$ 
is fitted with a sigmoid curve, as discussed 
in section~\ref{sec:simulations}.
The efficiency curves for each detector
are combined to get an overall triple-coincidence
efficiency, 
averaging over source direction and polarization,
assuming that the coincident efficiency is the product of the
individual efficiencies (as a function of amplitudes
at each detector).
The effects of our (very limited) post-coincidence
processing, including the choice of coincident 
time windows, clustering of multiple triggers from
a single detector,
and consistency of burst trigger parameters
from the three detectors, have been studied.
None of these studies reveal a significant source of error.
It should be noted here that future results
will employ far more detailed post-coincidence
processing (see section \ref{sec:future}),
including much tighter coincident time windows,
and these issues will be carefully re-evaluated.

By far the largest source of systematic error associated
with the efficiency determination is the uncertainty
in the detector response to gravitational waves
as obtained through the calibration 
procedure \cite{ref:calibration,ref:calib2}.
We have combined the estimated uncertainty in the DC calibration
with potential uncertainties in the frequency response,
convolved with the frequency spectra of our
modeled waveforms. We conservatively estimate
an overall systematic uncertainty of 20\%\ in the 
strain scale for our efficiency curves.
Our curves of 
upper-limit versus gravitational wave strain
(Fig.~\ref{fig:TFfinal})
reflect this uncertainty by assuming that the detectors'
response is 20\%\ less sensitive than the nominal value.

We assign no systematic error associated with
our choice of modeled waveforms since these 
are used to define the
set of bursts which are targeted by this search.

We assign no systematic errors associated with
our choice of data used 
or our BLRMS cut or calibration cut
(section \ref{sec:preselection}).
These procedures simply fix our observation time.

The upper limit on the number of observed bursts
depends on our estimate of the backgrounds,
as discussed in section \ref{sec:background}.
The statistical errors associated with these
estimations are marginalized over in the process
of establishing the confidence bands;
since these errors are small, this has a negligible effect 
on the resulting upper limits.
We have performed a variety of studies to 
search for and estimate sources of systematic errors
in the background estimate, including the time
dependence of the background rate, 
various time-lag procedures,
excluding anomalous data stretches, and other consistency checks.
No sources of additional systematic error
associated with the background rate
are found to be significant.



\section{Summary and discussion}\label{sec:summary}

We have searched for gravitational wave bursts of astrophysical origin
using data taken during the first Science Run of the three LIGO detectors.
A total of 35.5 hours of triple coincidence
observation time satisfied our data quality requirements.

We employed two different algorithms for the identification of 
candidate burst events in the gravitational wave data channel.
With the algorithm for which we chose to present a final result
(the \mycaps{TFCLUSTERS} event trigger generator),
we observe numbers of events that are reasonably consistent
with expectations for random accidental coincidences
of events originating from noise fluctuations
in the three detectors. 
We thus limit the excess event rate to be less than
1.6 per day , at 90\%\ confidence level.

We estimate our efficiency for the detection of
linearly polarized 
gravitational wave bursts incident on the detector array
with a range of amplitudes and 
averaged over source direction and wave polarization,
by injecting simulated signals into the raw S1 data streams
and performing the search as it is done on the raw data itself.
We focus on two simple, 
ad hoc waveform morphologies (section \ref{sec:waveforms}):
Gaussians with a range of durations, and sine-Gaussians
with a Q of 9, and a range of central frequencies.
With these, we evaluate the (waveform-dependent) detection 
efficiencies as a function of strain per root hertz, $h_{rss}$.
We then combine the rate limit with the efficiencies
to establish exclusion regions in the rate-versus-strength
plane; representative examples are shown in 
Fig.~\ref{fig:TFfinal}.
These constitute the results of the search reported in this paper.

\subsection{Comparison with other burst searches}\label{sec:comparison}

Searches for gravitational wave bursts have been a main focus of the
observational program of the resonant-mass detector community, and our
work was able to benefit from their prior work.  The most recent
analysis by the IGEC consortium \cite{ref:IGEC2003},
which includes data from five
detectors spread around the world, has presented its upper limits for
bursts in the form of a rate-amplitude diagram, in much the same style
as our Fig.~\ref{fig:TFfinal}.

The IGEC result (Fig.~13 of \cite{ref:IGEC2003})
bounds the rate for GW burst events 
with large amplitude 
to be less than $\sim 4\times 10^{-3}$ events per day.
This rate bound is much stronger than the ones reported here
due to the much longer observation time of the IGEC run.
The rate bound gets rapidly worse for lower amplitude
bursts, due to increasing background and decreasing
detection efficiency.

To crudely compare the sensitivity of the two searches 
to low-amplitude bursts, we can define a ``detection threshold''
as the value of the burst amplitude at which the rate limit 
is two times worse than the limit for large amplitude bursts.
For the analysis presented here, this is the
50\%\ efficiency point $h_{rss\, 1/2}$
reported in Table \ref{tbl:S1sens}.

One difference between the IGEC work and ours is that their
instruments have relatively narrow frequency bandwidths, 
and so are sensitive
to a different measure of burst strength.  Their detectors measure
the Fourier magnitude $|\tilde{h}(f_b)|$ of a signal
waveform at the bars' resonant frequency $f_b \approx 900$ Hz,
incident at optimal source direction and polarization.

However, if we consider a specific waveform with 
frequency content dominantly at or near
at $f_b$, such as our sine-Gaussians 
with central frequency $f_0 = 850$ Hz,
the bar sensitivities and 
the interferometric detectors' sensitivities can be directly compared, 
over the relatively narrow frequency band where bar detectors are
most sensitive.
The conversion from $|\tilde{h}(f_b)|$ to $h_{rss}$ for bars
for the sine-Gaussian family of signals (Eqn.~\ref{eq:SG})
is calculated to be

\begin{equation}
h_{rss} = \frac{|\tilde{h}(f_b)|}{\sqrt{\tau}}   
\frac{(\frac{2}{\pi})^{\frac{1}{4}}}
{( e^{-\pi^2 \tau^2 (f_b - f_0)^2} - e^{-\pi^2 \tau^2 (f_b + f_0)^2} )} .
\label{eq:bars}
\end{equation}

We focus on sine-Gaussians
with central frequency $f_0 = 850$ Hz,
incident at optimal direction and polarization.
Using Fig.~13 of \cite{ref:IGEC2003} and Eqn.~\ref{eq:bars},
the IGEC detection threshold is 
roughly $h_{rss} \approx 1\times 10^{-19}$ Hz$^{-1/2}$.
To compare this with our sensitivity for
850 Hz sine-Gaussian bursts 
($h_{rss\, 1/2} = 7.3\times 10^{-19}$ Hz$^{-1/2}$,
Table \ref{tbl:S1sens}),
we must first correct for our 
averaging over direction and polarization
(section \ref{sec:direction}).
This yields an
amplitude at 50\%\ efficiency for waves with
optimal orientation of 
$h_{rss} \approx 2.6\times 10^{-19}$ Hz$^{-1/2}$.
Note that this detection threshold for LIGO 
is established by determining the loss of efficiency
for fixed threshold,
while for IGEC it is established by observing 
an increase in background events as the threshold is varied.
Nonetheless, this measure of detection threshold 
permits a rough comparison of the search sensitivites \cite{ref:IGECcom},
and we see that the IGEC search \cite{ref:IGEC2003}
has a somewhat greater sensitivity to 850 Hz sine-Gaussian bursts
than the one presented here.

For all other waveforms shown in Fig.~\ref{fig:TFfinal},
and for other waveforms with significant spectral amplitude in
a broad range of frequencies away from $f_b$,
the LIGO constraints are more stringent than the IGEC results, 
due to the broad band response
of the interferometric detectors. 
For sine gaussians at 554 Hz, the ratio of peak
spectral density of the pulse to the spectral density
in the resonant-mass detector band is less than 10\%\ 
for NIOBE and negligible in the other four IGEC detectors. 
For sine gaussians at 1304 Hz,
ALLEGRO, AURIGA, EXPLORER and NAUTILUS receive 
spectral densities in their bands
that are only a few percent of peak spectral density for the pulse, 
with negligible spectral density in NIOBE's band.
The resonant mass detectors also receive relatively small
spectral density in their bands from 
gaussian waveforms compared to the LIGO detectors, 
unless $\tau$ is less than 1 ms.
This emphasises the importance of broad band sensitivity 
in searching for unmodeled gravitational-wave bursts. 
Ongoing work to broaden the response of resonant mass detectors
should improve sensitivity to other waveforms in the future.

The only previously published results on searches for burst
events with broadband interferometric detectors that we are aware of
are in Ref.~\cite{ref:UGMPQ} (but see also Ref.~\cite{ref:Forward}).
In Ref.~\cite{ref:UGMPQ}, prototype detectors developed
by the University of Glasgow and Max Planck Institute for Quantum
Optics were operated for an effective coincident observing
period of 62 hours in 1989.
They searched for bursts with significant frequency content
in the band from 800 to 1250 Hz.
They considered the waveform 
$h(t) = h_{peak} \sin(2\pi f_m t) / (2\pi f_m t)$ \cite{ref:Bernie},
which has constant Fourier magnitude from 0 to $f_m = 1250$ Hz,
and in-band ($\Delta f = 1250-800 = 450$ Hz) root-sum-square amplitude
$h_{rss} = \sqrt{2\Delta f}/(2 f_m) h_{peak}$.
They observe no events with $h > h_{peak} = 4.9\times 10^{-16}$,
or $h_{rss} = 5.9\times 10^{-18} \,{\rm Hz}^{-1/2}$,
averaging over wave polarizations and incident directions. 
Therefore, they set an upper limit on the rate of bursts
with strain greater than this value,
of 0.94/day. 
Their sensitivity can be compared with the strain sensitivities
reported here at 50\%\ efficiency, for sine-Gaussians
with central frequencies of 850 Hz and 1304 Hz:
$h_{rss\, 1/2} = 7.3\times 10^{-19} \,{\rm Hz}^{-1/2}$ and
$h_{rss\, 1/2} = 1.4\times 10^{-18} \,{\rm Hz}^{-1/2}$, respectively
(Table~\ref{tbl:S1sens}).

\subsection{Directions for improved analysis in the future}\label{sec:future}

LIGO's second science run (S2) accumulated data for 8 weeks in early 2003.
At most frequencies, the noise in the three LIGO detectors was improved
compared to the noise level of the S1 data presented here
by a factor of 10. Some improvements in the
stability of the noise were also achieved.  The in-lock duty cycles of the
detectors were comparable to those obtained during S1, 
but tighter monitoring of
the detectors' noise levels and calibration should lead to significantly
less loss of data than was suffered in S1. Even without improvements in our
analysis methodology, we expect to obtain results from the S2 data that are
an order of magnitude more sensitive in amplitude, and observation times
that are increased by at least a factor of four 
over the results presented here.

Based on lessons learned during the S1 analysis, we are preparing
numerous
improvements and additions to our search methodology for the S2 data set.
The pipeline
presented here can be improved with more attention to optimizing and
characterizing our event trigger generators. Obvious areas for improvement
are better prefiltering, and better time resolution for both \mycaps{SLOPE} and
\mycaps{TFCLUSTERS}. We have also implemented an adaptive threshold for the \mycaps{SLOPE}
event trigger generator to make its event rate less sensitive to variations
in detector noise.  As the detector performance becomes more stable and
closer to the design sensitivity, safe and effective vetoes based on auxiliary
channels that monitor the environment and interferometer sensing and
control will be applied to reduce the number of spurious event
triggers.  New event trigger generators, using a variety of detection
techniques ({\it e.g.}, that proposed in \cite{ref:power}), will be tested.

After initial (coarse) identification of coincident events with 
improved versions of the event
trigger generators, the gravitational wave data channel time series can be
reexamined to further reduce the background of accidental
coincidences.  Cross-correlation of the gravitational wave channels from
multiple detectors can tighten our coincidence window so that it is limited
only by the light-travel time between detector sites, 
and test whether the
event amplitudes and waveforms are consistent with the common origin of a
gravitational plane wave.

Future searches will include a more astrophysical style of interpretation, 
setting limits on populations of events in three-dimensional space. 
Efficiency simulations will include more realistic waveforms, such as black 
hole ringdowns or supernova waveforms \cite{ref:ZM,ref:DFM}.  
Higher sensitivity for 
modeled bursts can be obtained using matched filter techniques.  Longer 
runs will give more opportunities to search for gravitational wave bursts 
coincident with gamma ray burst events, using the methodology described in 
\cite{ref:FinnGRB}.

Finally, and crucially, we are developing criteria by which we can 
establish confidence in the detection of gravitational wave bursts both 
statistically and as a single large amplitude burst event. 
For single burst event candidates, we will use
information from all available detectors to reconstruct our best estimates 
of the gravitational wave direction, polarization, and waveform.

\begin{acknowledgments}
The authors gratefully acknowledge the support of the United States National
Science Foundation for the construction and operation of the LIGO Laboratory and
the Particle Physics and Astronomy Research Council of the United Kingdom, the
Max-Planck-Society, and the State of Niedersachsen/Germany for support of the
construction and operation of the GEO600 detector. The authors also gratefully
acknowledge the support of the research by these agencies and by the Australian
Research Council, the Natural Sciences and Engineering Research Council of
Canada, the Council of Scientific and Industrial Research of India, the
Department of Science and Technology of India, the Spanish Ministerio de Ciencia
y Tecnologia, the John Simon Guggenheim Foundation, the David and Lucile Packard
Foundation, the Research Corporation, and the Alfred P. Sloan Foundation. 
This document has been assigned LIGO Laboratory document number 
LIGO-\ligodoc.
\end{acknowledgments}

\bibliographystyle{apsrev}

\begin{thebibliography}{46}
\bibitem[{ref({\natexlab{e}})}]{ref:meaculpa}
\bibinfo{note}{This is an instance where our parameter tuning, event property
  estimation, or other analysis procedure was inadequately optimized prior to
  being frozen for the analysis of the full S1 data set. See the discussions at
  the beginning of section~\ref{sec:pipeline} and in section~\ref{sec:future}.}
\end{thebibliography}

\end{document}